%% file: main.tex
\documentclass[letterpaper]{article} % DO NOT CHANGE THIS
\usepackage[draft]{aaai2026}
\usepackage{times}  % DO NOT CHANGE THIS
\usepackage{helvet}  % DO NOT CHANGE THIS
\usepackage{courier}  % DO NOT CHANGE THIS
\usepackage[hyphens]{url}  % DO NOT CHANGE THIS
\usepackage{graphicx} % DO NOT CHANGE THIS
\usepackage{natbib}  % DO NOT CHANGE THIS AND DO NOT ADD ANY OPTIONS TO IT
\usepackage{caption} % DO NOT CHANGE THIS AND DO NOT ADD ANY OPTIONS TO IT
\usepackage{booktabs}
\usepackage{array}
\usepackage{algorithm}
\usepackage{algorithmic}

\usepackage{newfloat}
\usepackage{listings}
\usepackage[utf8]{inputenc}
\usepackage{lipsum}
\usepackage{microtype}
\usepackage{amsmath}
\usepackage{amsfonts}
\usepackage{xcolor}
\usepackage{multirow}
\usepackage[T1]{fontenc}
\definecolor{c4}{RGB}{255,225,187}
\definecolor{c2}{RGB}{209, 233, 184}
\definecolor{c3}{RGB}{218,243,246}
\definecolor{c1}{RGB}{249, 229, 229}
\definecolor{c5}{RGB}{255, 128, 128}
\definecolor{c6}{RGB}{251, 132, 002}

\usepackage{listings}
\usepackage[most]{tcolorbox}
\tcbuselibrary{listings}

\tcbset{
  listing engine=listings,
  colback=white,           % 背景色
  colframe=black,       % 柔和边框色 (50% 黑)
  colbacktitle=white,
  coltitle=black,
  fonttitle=\bfseries,
  boxrule=0.5pt,           % 边框粗细
  arc=3pt,                 % 圆角半径
  outer arc=3pt,
  listing only,
  breakable,               % 可分页
  enhanced jigsaw,
}

\DeclareCaptionStyle{ruled}{labelfont=normalfont,labelsep=colon,strut=off} % DO NOT CHANGE THIS
\floatstyle{ruled}
\newfloat{listing}{tb}{lst}{}
\floatname{listing}{Listing}

\title{M2IO-R1: An Efficient RL-Enhanced Reasoning Framework for Multimodal Retrieval Augmented Multimodal Generation}

\author{
    Zhiyou Xiao\textsuperscript{\rm 1}\equalcontrib, Qinhan Yu\textsuperscript{\rm 1}\equalcontrib\thanks{Work done during an internship at Huawei.}, Binghui Li\textsuperscript{\rm 1}\equalcontrib, Geng Chen\textsuperscript{\rm 2}, Chong Chen\textsuperscript{\rm 3}\thanks{Corresponding authors.}, Wentao Zhang\textsuperscript{\rm 1$\ddagger$}
}
\affiliations{
    \textsuperscript{\rm 1}Peking University, \textsuperscript{\rm 2}Jilin University, \textsuperscript{\rm 3}Huawei Cloud BU\\
    \{xiaozhiyou, yuqinhan\}@stu.pku.edu.cn, \{libinghui, wentao.zhang\}@pku.edu.cn\\ chengeng0201@gmail.com, chenchong55@huawei.com
}

\usepackage{bibentry}
\begin{document}

\maketitle

\begin{abstract}
Current research on Multimodal Retrieval-Augmented Generation (MRAG) enables diverse multimodal inputs but remains limited to single-modality outputs, restricting expressive capacity and practical utility.
In contrast, real-world applications often demand both multimodal inputs and multimodal outputs for effective communication and grounded reasoning.
Motivated by the recent success of Reinforcement Learning (RL) in complex reasoning tasks for Large Language Models (LLMs), we adopt RL as a principled and effective paradigm to address the multi-step, outcome-driven challenges inherent in multimodal output generation.
Here, we introduce M2IO-R1, a novel framework for Multimodal Retrieval-Augmented Multimodal Generation (MRAMG) that supports both multimodal inputs and outputs.
Central to our framework is an RL-based inserter,  Inserter-R1-3B, trained with Group Relative Policy Optimization to guide image selection and placement in a controllable and semantically aligned manner.
Empirical results show that our lightweight 3B inserter achieves strong reasoning capabilities with significantly reduced latency, outperforming  baselines in both quality and efficiency. 
\end{abstract}

\input{Intro}

\input{Related_Work}

\input{Method}

\input{Experiments}

\input{Conclusion}

\newpage
\bibliography{aaai2026}

 \newpage
 \appendix
 \input{Appendix}

%\bibliography{aaai2026}

% \newpage
% \input{ReproducibilityChecklist}

\end{document}

%% file: Intro.tex
\section{Introduction}
Retrieval-Augmented Generation (RAG) enhances the capabilities of Large Language Models (LLMs) by incorporating external knowledge, thereby improving factual accuracy and mitigating hallucinations \citep{lewis2020retrieval,zhao2024retrieval}. 
Recent advances of Multimodal Large Language Models (MLLMs) \citep{team2023gemini,yin2024survey} have extended RAG into the multimodal domain, enabling joint reasoning over textual and visual inputs.
However, existing Multimodal RAG frameworks predominantly generate text-only outputs \citep{chen2022murag,liu2023learning}, which limits their effectiveness in real-world scenarios where visual content is crucial for interpretability, grounding, or user understanding. 

\begin{figure}[t]
    \centering    \includegraphics[width=0.99\linewidth]{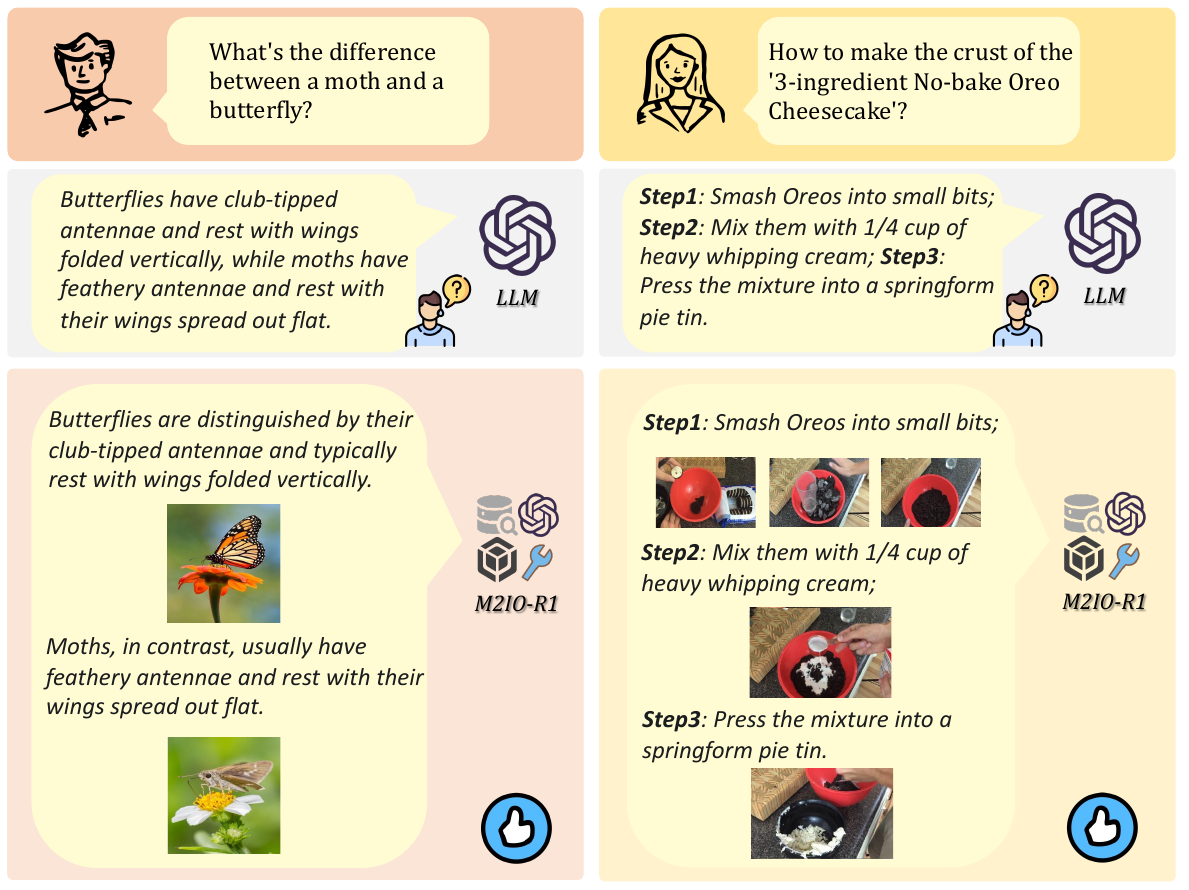}
   \caption{ 
    Illustration of two practical scenarios where users query for multimodal responses. The middle row illustrates plain LLM outputs, whereas the bottom row depicts the expected multimodal outputs.
    }
    \label{fig:example}
\end{figure}

This limitation becomes particularly pronounced in applications where visual content is central to interpretation and usability \citep{zhu2024murar,yu2025mramg}. For instance, visual modalities significantly enhance user engagement in domains such as travel recommendations and product descriptions. 
In educational or instructional scenarios, however, visual information is not merely helpful—it is critical for ensuring clarity and safety.
As illustrated in Figure~\ref{fig:example}, step-by-step visual guides are indispensable for procedural tasks like distinguishing moths from butterflies or preparing a recipe, as they reduce ambiguity and mitigate operational risks. 
This growing reliance on visual support underscores a broader shift in user expectations toward intuitive, multimodal communication. 
Consequently, text‑only RAG systems face practical limitations, rendering multimodal output generation essential for real‑world applications.

Research on enabling multimodal output can broadly be categorized into two paradigms. The first, \textbf{interleaved multimodal generation}, employs autoregressive models \citep{sun2024generative, li2024autoregressive,tian2024visual} or diffusion models \citep{peebles2023scalable, bao2022analytic} trained on paired image-text data to jointly produce textual and visual content \citep{tian2024mm}. 
While effective for tasks like storytelling and webpage synthesis, they face two key challenges: (1) difficulty maintaining coherent narrative flow and entity grounding, often resulting in hallucinations and semantic inconsistency \citep{chen2024interleaved}; and (2) limited real-world applicability, as users typically prefer visuals from authentic sources (e.g., photographs, schematics, scientific figures) to fully AI-generated content \citep{bui2024ai}.
The second paradigm, \textbf{Multimodal Retrieval-Augmented Multimodal Generation (MRAMG)}, emphasizes factuality by integrating retrieved real-world image-text content into the generation process.
Rather than synthesizing visuals from scratch, MRAMG leverages authentic visuals to enhance semantic alignment and grounded reasoning.
Despite its practical requirements, this paradigm remains underexplored. Most existing efforts focus on constructing evaluation benchmarks and typically adopt training-free methods that rely on LLM-prompt design or heuristic image insertion strategies \citep{zhu2024murar,ma2024multi,yu2025mramg}.

In this work, we focus on the second paradigm—MRAMG. Given its potential to enhance factuality, user trust, and applicability in real-world scenarios, we aim to systematically investigate and advance this direction. Crucially, the core challenge in MRAMG lies in deciding which images to select, and where and in what order to insert them—\textit{a multi-step, outcome-driven process} that conventional supervised learning struggles to capture \citep{ma2024multi}. This framing of image selection and placement as a sequential decision problem makes Reinforcement Learning (RL) a particularly suitable approach. 
Specifically, inspired by the recent success of RL in optimizing LLMs for other complex reasoning tasks \citep{jaech2024openai,guo2025deepseek,jin2025search,wang2025vrag}, we propose M2IO-R1: a lightweight yet effective framework for MRAMG with \underline{\textbf{M}}ulti\underline{\textbf{M}}odal \underline{\textbf{I}}nput-\underline{\textbf{O}}utput capabilities. The framework follows a four-stage pipeline: (1) retrieve relevant text and images; (2) generate a textual response; (3) apply an RL-based inserter to select and place visual content; and (4) interleave text and images into a coherent, grounded output. 

Our framework adopts a decomposed strategy: it first generates text and then inserts images, reducing complexity and improving controllability. Central to this design is an RL-based inserter—an extension of the DeepSeek-R1 paradigm \citep{guo2025deepseek}—which is trained via  GRPO \citep{shao2024deepseekmath} to make outcome-driven decisions on illustration selection and placement.
 Notably, the 3B inserter ``\textit{punches above its weight},'' exhibiting strong reasoning capabilities despite its significantly lower inference cost and latency.
This result highlights the broader applicability of RL to general-purpose multimodal tasks beyond math and code.

Furthermore, our study presents a comprehensive investigation into non-generative multimodal outputs—an important yet underexplored direction within multimodal understanding. We hope our findings will encourage further exploration and accelerate the development of reliable, interpretable, and practically deployable multimodal systems.

In summary, our main contributions are as follows:
\begin{itemize}
    \item
    We introduce M2IO-R1, a novel framework for the emerging MRAMG task that decomposes the task into four stages—retrieval, generation, illustration, and merging—enabling controllable and interpretable multimodal outputs tailored to the unique challenges of MRAMG.
    \item
    We develop an RL‑enhanced image inserter, which is trained using GRPO-based RL framework to address MRAMG's core challenge: selecting appropriate images and placing them optimally within text—a multi‑step, outcome‑driven process that conventional supervised learning struggles to capture.
\item
    We conduct extensive empirical evaluations, showing that our method ``punches above its weight''—delivering competitive or superior performance while significantly reducing computational cost and latency.
\end{itemize}

From a technical perspective, recent advances such as DeepSeek-R1 \citep{guo2025deepseek} demonstrate that discarding process rewards in favor of outcome-based supervision can deliver substantial improvements. 
Motivated by this, we identify a multimodal scenario where outcome rewards can be precisely computed. By restricting the inserter's output to image indices and insertion positions, we design a simple yet interpretable rule‑based reward, marking a notable step forward in multimodal reasoning.

%% file: Related_Work.tex
\section{Related Work}
\subsection{Multimodal RAG}
\begin{figure*}[t]
    \centering    \includegraphics[width=0.99\linewidth]{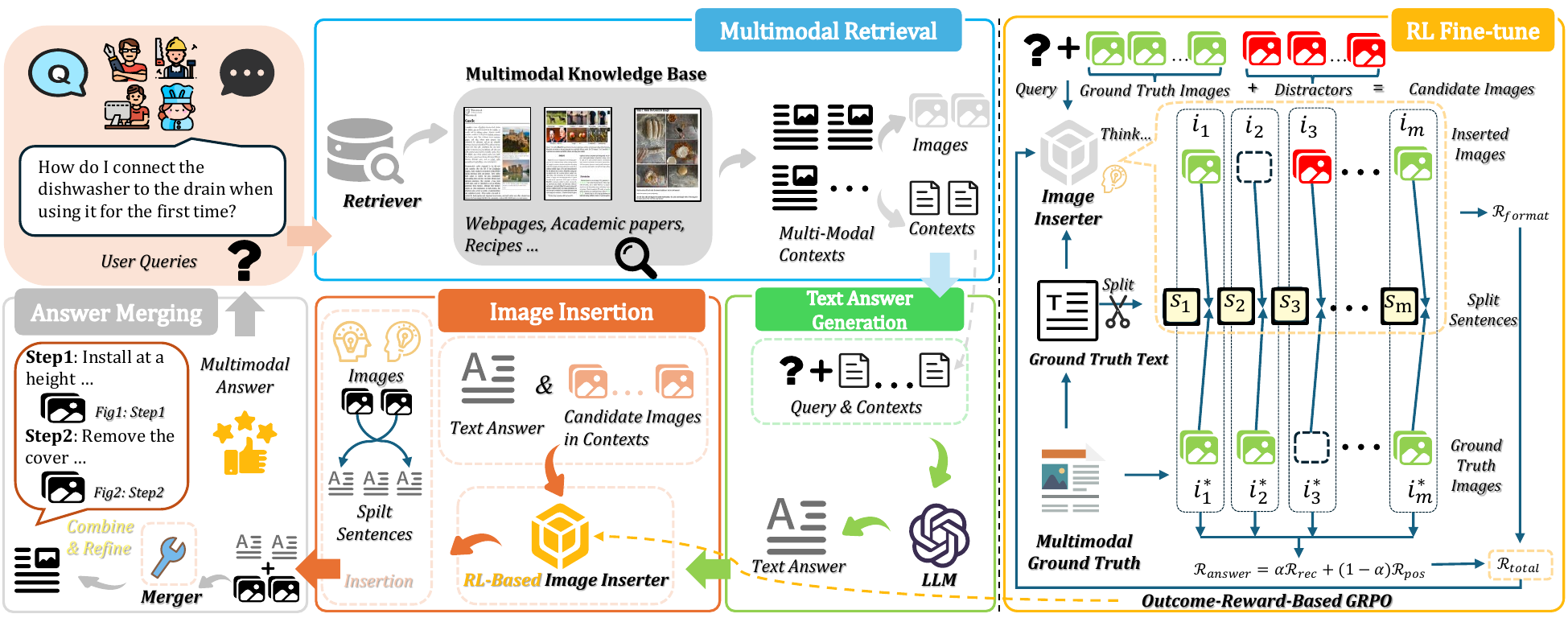}
   \caption{
Overview of the M2IO-R1 framework. Left : The pipeline consists of four sequential stages—(1) multimodal retrieval, (2) text answer generation, (3) image insertion, and (4) answer merging—to produce coherent and refined multimodal answers. Right : Training process of the RL-based insertion module (Inserter-R1), optimized with GRPO.
    }
    \label{fig:main_pip}
\end{figure*}

Retrieval-Augmented Generation (RAG) enables Large Language Models (LLMs) to access accurate and up-to-date external knowledge, thereby improving response accuracy \citep{hallucination, zhao2024retrieval}. While traditional RAG frameworks primarily rely on LLM-based agents to retrieve textual information via search engines \citep{self_rag, chen2024agent}, recent advancements in Multimodal Large Language Models (MLLMs) have extended RAG to support cross-modal retrieval and reasoning over both textual and visual modalities \citep{yu2024visrag}.

Despite these advances, most existing methods focus on processing multi-modal inputs, while largely neglecting the challenge of modality augmentation in generated outputs \citep{mei2025survey}. Recently, the task of Multi-modal Retrieval-Augmented Multi-modal Generation (MRAMG) has gained attention \citep{zhu2024murar, yu2025mramg}, aiming to generate interleaved text-image outputs. To address this task, MRAMG-Bench \citep{yu2025mramg} formulates image insertion as a bipartite graph matching problem using heuristic strategies, whereas M2RAG \citep{ma2024multi} adopts a prompt-based  pipeline to guide image placement and textual refinement. However, these methods primarily focus on dataset construction and inference-time strategies, without exploring trainable architectures or optimization techniques for MRAMG. To bridge this gap, we propose a lightweight, flexible, and RL-based framework that jointly optimizes the generation of multi-modal outputs.

%\vspace{-0.5em}
\subsection{RL-Enhanced Large Reasoning Models} Recent advances in Large Reasoning Models (LRMs), such as OpenAI-o1 \citep{jaech2024openai} and DeepSeek-R1 \citep{guo2025deepseek}, demonstrate that reinforcement learning (RL) can significantly improve the multi-step reasoning capabilities of LLMs. Building upon these findings, recent efforts have extended RL techniques to MLLMs to enhance their reasoning performance \citep{meng2025mm, liu2025visual}. For instance, Vision-R1 \citep{huang2025vision} applies RL with GRPO to structured textual representations derived from visual inputs, thereby improving multimodal reasoning. Similarly, LMM-R1 \citep{peng2025lmm} adopts a two-stage RL framework that transfers reasoning skills learned in text-only tasks to multimodal domains. 

Despite these promising developments, existing RL applications in large models remain largely concentrated in narrow domains such as code generation, mathematical problem solving, and multi-hop question answering \citep{zhou2025reinforced, wang2025vrag}, with limited exploration of general-purpose reasoning capabilities required for real-world scenarios. In contrast, our work extends R1-style reinforcement learning beyond these conventional domains, proposing a unified and lightweight framework that enhances multimodal reasoning and supports interleaved text-image generation.

%% file: Method.tex
\section{Method}
In this section, we first describe the data curation pipeline, then outline the overall M2IO architecture, and finally present an RL-based image inserter.

\subsection{Data Preparation}
\label{sub_data_se}

To facilitate RL training with high-quality supervision signals amid the scarcity of interleaved text-image datasets with precise image placement annotations, we construct the M2IO-Inserter dataset by systematically sampling from six distinct subsets of the MRAMG-Bench benchmark \citep{yu2025mramg}.

\paragraph{Annotation and Diversity.} Our construction process emphasizes fine-grained image-text alignment annotations, which are essential for designing effective reward functions. These annotations enable RL agents to learn optimal image insertion strategies in complex multimodal contexts. The dataset encompasses three representative categories—web content, academic materials, and instructional documents—covering a diverse range of real-world multimodal scenarios. Each sample undergoes comprehensive manual inspection and annotation to ensure both the accuracy of image placement and data integrity.

\paragraph{Difficulty Stratification.} To ensure both quality and diversity, we implement a tri-model evaluation framework (GPT-4o \citep{gpt4o}, Claude \citep{anthropic_claude_2024}, Gemini \citep{team2024gemini}) that independently assesses sample difficulty through answer accuracy scoring. The difficulty scores from three models are normalized, averaged, and stratified using tercile thresholds to categorize each sample into one of three levels: easy, medium, or hard.

\paragraph{Sampling Protocols.} For each query $q$  , we include all relevant positive images ($ \mathcal{I}^{+} $), and select an equal number of negative distractor images ($ \mathcal{I}^{-} $), 
comprising both random and hard distractors retrieved automatically, thereby enforcing a strict $ 1\!:\!1 $ positive-to-negative ratio.
Following filtering and difficulty-balanced sampling, we adopt two distinct data construction protocols tailored to different evaluation goals:
\begin{itemize}
    \item \textbf{Full-source split}: 
    Randomly sampled from all six MRAMG-Bench datasets, divided equally into 50\% training and 50\% evaluation. The training set comprises 2.4k examples with a uniform difficulty distribution.
    \item \textbf{Web-focused split}: We focus on three web-related datasets—\texttt{Wit}, \texttt{Web}, and \texttt{Wiki}—using 80\% of their samples for training (1.4k examples) and 20\% for testing. Samples from the other three datasets are reserved exclusively for evaluation. This split enables testing generalization beyond the web domain.  
\end{itemize}

\subsection{Pipeline of M2IO-R1}
We propose MIMO-R1, a novel MRAMG framework that decomposes the overall generation process into four modules: a retriever, a text generator, an image inserter, and a merger. This modular design enables controllable and flexible multimodal answer generation by seamlessly integrating textual and visual information.

\subsubsection{Retriever} serves as the foundational stage, responsible for identifying and retrieving the most contextually relevant multimodal documents from the knowledge base $\mathcal{D} = \{d_1, d_2, \dots, d_n\}$, given a user query $q$. Specifically, we employ an embedding model $\mathcal{E}(\cdot)$ that projects queries and documents into a continuous embedding space. The semantic relevance between the query $q$ and a document $d$ is quantified by computing the cosine similarity between their respective embeddings:
\begin{equation}
    \operatorname{sim}(q, d) = \frac{\mathcal{E}(q)^\top \mathcal{E}(d)}{\|\mathcal{E}(q)\|\|\mathcal{E}(d)\|}.
\end{equation}
% $
% \operatorname{sim}(q, d) = \frac{\mathcal{E}(q)^\top \mathcal{E}(d)}{\|\mathcal{E}(q)\|\|\mathcal{E}(d)\|}.
% $

For each query $q$, the retriever returns the top-$k$ most semantically relevant multimodal documents, denoted as $\mathcal{D}^*_q = \{d_{j_1}, d_{j_2}, \dots, d_{j_k}\}$.

\subsubsection{Text Generator}

Following the retrieval stage, the text generator module is responsible for producing a coherent and informative textual answer grounded in the retrieved documents. Specifically, we utilize a large generative model $\mathcal{G}$ to generate the textual answer $\mathcal{A}_{\text{txt}}$, formally defined as
\begin{equation}
    \mathcal{A}_{\text{txt}} = \mathcal{G}(\mathcal{P}_{\text{txt}}, q, \mathcal{T}_q),
\end{equation}
where $\mathcal{P}_{\text{txt}}$ denotes the textual answer generation prompt, and ${\mathcal{T}_q} \subseteq \mathcal{D}^*_q$ represents the set of textual segments extracted from the retrieved multimodal documents $\mathcal{D}^*_q$ that are most relevant to the query $q$.

\subsubsection{Image Inserter} 

After generating the textual answer, we split the answer $\mathcal{A}_{\text{txt}}$ into sentences, obtaining $\mathcal{S} = \{s_1, s_2, \dots, s_m\}$. Then, we invoke the image inserter to insert images. For each sentence, the inserter selects at most one image from the retrieved image set $\mathcal{I}_q$, or chooses to insert no image at all. This results in the following image-augmented textual answer:
\begin{equation}
\mathcal{A}_{\text{img}} = \mathcal{M}(\mathcal{P}_{\text{insert}},q, \mathcal{S}, {\mathcal{I}_q}),
\end{equation}
where $\mathcal{P}_{\text{insert}}$ denotes the image insertion prompt, and $\mathcal{M}$ is the image insertion function that associates images with their corresponding sentences based on the query $q$ and the retrieved image set $\mathcal{I}_q$.

\subsubsection{Merger}

The merger module is responsible for combining the original textual answer $\mathcal{A}_{\text{txt}}$ and the image-augmented answer $\mathcal{A}_{\text{img}}$. This is achieved by integrating both the text and images into a single, coherent multimodal output. The final multimodal output is produced as:
\begin{equation}
    \mathcal{A}_{\text{final}} = \operatorname{Merge}(\mathcal{A}_{\text{txt}}, \mathcal{A}_{\text{img}}),
\end{equation}
where the merging function $\operatorname{Merge} (\cdot,\cdot)$ integrates the text and images, ensuring that the images are correctly placed within the generated text to create a seamless, informative, and visually enriched response.

\subsection{RL-Based Image Inserter}
Inspired by DeepSeek-R1 \citep{guo2025deepseek}, we adopt a RL approach to improve the inserter's performance. We conceptualize illustration—particularly multi-image illustration—as a reasoning task, wherein the inserter is expected to generate intermediate reasoning steps that support accurate image placement. To this end, we employ the GRPO algorithm \citep{shao2024deepseekmath}, with a tailored reward function design as detailed below.

\subsubsection{Think Process}
To encourage the model to output its reasoning process during image selection and placement, we design a structured prompt that guides it through the necessary steps. The prompt first requests an analysis of the relevance of candidate images to the input text and their appropriate insertion positions.

Additionally, the model is explicitly instructed to generate an intermediate reasoning trace, enclosed within \textit{<think>} tags, where it justifies its decisions based on image content, textual relevance, and optimal placement. This reasoning output aims to enhance the transparency of the decision process and ensure coherent multimodal integration. The final selected images and their corresponding positions are then returned within \textit{<answer>} tags, formatted as a dictionary with image IDs as keys and sentence indices as values. This design promotes both the interpretability and effectiveness of the model, particularly in multi-image scenarios involving complex image-text relationships.

\subsubsection{Reward Design}
We design a rule-based reward function composed of two components: a format reward and an answer reward. The format reward evaluates whether the model's output adheres to the required structural constraints:
\begin{equation}
    \mathcal{R}_{\text{format}} = \begin{cases}
        0, &\text{if the format is incorrect;}
        \\
        1, &\text{if the format is correct.}
    \end{cases}
\end{equation}
The answer reward is computed only for validly formatted outputs and considers both the correctness of image selection and placement. We define the image insertion result as 
$\mathcal{I}=\{i_1,i_2,\dots,i_m\}$, where each element $i_j$ corresponds to the image inserted after sentence $s_j$. An element $i_j = \emptyset$ signifies that no image was inserted. Let $\mathcal{I}^*=\{i_1^*,i_2^*,\dots,i_m^*\}$ denote the ground-truth image sequence. We then define image recall as 
\begin{equation}
    \mathcal{R}_{\text{rec}} = \frac{|\mathcal{I}\cap \mathcal{I}^*|}{|\mathcal{I}^*|},
\end{equation}
%$\mathcal{R}_{\text{rec}} = {|\mathcal{I}\cap \mathcal{I}^*|}/{|\mathcal{I}^*|}$ 
and position accuracy as 
\begin{equation}
    \mathcal{R}_{\text{pos}} = \frac{1}{m}\sum_{k=1}^{m}\mathbb{I}\{i_k = i_k^{*}\}.
\end{equation}
%$\mathcal{R}_{\text{pos}} = \frac{1}{m}\sum_{k=1}^{m}\mathbb{I}\{i_k = i_k^{*}\}$.
The combined answer reward is computed as:
\begin{equation}
    \mathcal{R}_{\text{answer}} = \begin{cases} 0,&\text{if } \mathcal{R}_{\text{format}} = 0;\\
        \alpha \mathcal{R}_{\text{rec}} + (1-\alpha) \mathcal{R}_{\text{pos}},&\text{otherwise.}
    \end{cases}
\end{equation}
Finally, the total reward used for training is:
\begin{equation}
    \mathcal{R}_{\text{total}} = \mathcal{R}_{\text{format}} + \mathcal{R}_{\text{answer}}.
\end{equation}
This reward ensures valid outputs with accurate selections and precise placements in multimodal contexts.

%% file: Experiments.tex
\section{Experiments}
\label{sec:exper}

\subsection{Experiment Settings}
\paragraph{Benchmarks}
We evaluate our framework on three multimodal input-output benchmarks: MRAMG-Bench \citep{yu2025mramg}, the development sets of M2RAG \citep{ma2024multi}, and FTII-Bench \citep{ruan2024ftii}. 
The first two are MRAMG datasets, while FTII-Bench is a multimodal QA benchmark with streaming image illustrations.
These benchmarks span diverse domains and provide a comprehensive testbed for evaluating the M2IO framework on MRAMG.

\paragraph{Metrics}
For evaluation, we use six metrics: 
\begin{itemize}
    \item
    \textbf{Recall (Rec)} measures the proportion of ground‑truth images correctly included in the multimodal answer.
    \item 
    \textbf{F1-Score (F1)} measures the fidelity of the images in the multi-modal answer compared to the ground truth.
    \item 
    \textbf{Order Score (Ord) \citep{yu2025mramg}} measures the consistency of image order between the multimodal answer and the ground truth using a weighted edit distance.
    \item 
    \textbf{Relevance (Rel) \citep{zhu2024murar,ma2024multi,yu2025mramg}} assesses the semantic alignment between the visual content of the images and the surrounding query-answer text.
    \item 
    \textbf{Position Score (Pos)} evaluates the coherence of image placement within the generated multimodal answer.
    \item 
    \textbf{Overall Score (Ovr)} assesses the overall quality and helpfulness of a multimodal answer compared to the ground truth.
\end{itemize}
% \begin{itemize}
%     \item 
%     \textbf{Recall (Rec):} proportion of ground-truth (GT) images correctly included.
%     \item 
%     \textbf{F1 Score (F1):} the harmonic mean of precision (proportion of selected images matching the GT) and recall, providing a balanced evaluation of image quality in multimodal answers.
%     \item 
%     \textbf{Order Score (Ord) \citep{yu2025mramg}:} consistency with GT image order, measured by weighted edit distance.
%     \item 
%     \textbf{Relevance (Rel) \citep{zhu2024murar,ma2024multi,yu2025mramg}:} semantic alignment of images with query–answer text.
%     \item 
%     \textbf{Position Score (Pos):} coherence of image placement.
%     \item 
%     \textbf{Overall Score (Ovr):} overall quality and helpfulness relative to ground truth.
% \end{itemize}
Details are provided in the appendix. 

Referring to the availability of ground truth and prior work \citep{yu2025mramg, ma2024multi}, we adopt dataset‑specific metrics: Rec, F1, Ord, Rel, and Ovr on MRAMG‑Bench; F1 and Pos on FTII‑Bench; and Rel and Pos on M2RAG.
% For evaluation, we use six metrics: 
% (1) Recall (Rec) : proportion of ground-truth (GT) images correctly included; (2) F1 : the harmonic mean of precision (proportion of selected images matching the GT) and recall, providing a balanced evaluation of image quality in multimodal answers; (3) Order (Ord) :consistency with GT image order, measured by weighted edit distance; (4) Relevance (Rel) : semantic alignment of images with query–answer text; (5) Position Score (Pos) : coherence of image placement; 6) Overall Score (Ovr) : overall quality and helpfulness relative to ground truth. Details are provided in the appendix.
% Referring to the availability of ground truth and prior work \citep{yu2025mramg, ma2024multi}, we adopt dataset‑specific metrics: Rec, F1, Ord, Rel, and Com on MRAMG‑Bench; F1 and Pos on FTII‑Bench; and Rel and Pos on M2RAG.
\paragraph{Baselines}
We evaluate the two following  baselines:
\begin{itemize}
    \item \textbf{Single-Shot Strategy} \citep{ma2024multi,yu2025mramg}, which generates the complete multimodal response in a single pass based on the provided multimodal document.
    \item \textbf{Rule-Based Strategy} \citep{yu2025mramg} formulates image insertion as a weighted bipartite graph matching task by linking sentences and candidate images based on textual and semantic similarity.
\end{itemize}

M2IO includes three variants, distinguished by image inserter: M2IO-Base (training-free), M2IO-SFT (supervised fine-tuned), and M2IO-R1 (RL-based fine-tuned).

\begin{table*}[htbp]
  \centering
  % \small
  \setlength{\tabcolsep}{4.5pt}
  \caption{Performance comparisons between M2IO framework and the baselines on the MRAMG-Bench datasets. \textbf{T-Generator} denotes the text answer generator. The best score is in \textbf{bold} and the second best is \underline{underlined}.}
    \begin{tabular}{c|l|cccc|ccccc|cccc}
    \toprule
    \multirow{2}[2]{*}{\textbf{T-Generator}} & \multicolumn{1}{c|}{\multirow{2}[2]{*}{\textbf{Strategy}}} & \multicolumn{4}{c|}{\textbf{Arxiv}} & \multicolumn{5}{c|}{\textbf{Manual}}  & \multicolumn{4}{c}{\textbf{Web}} \\
          &       & \textbf{Rec} & \textbf{F1} & \textbf{Rel} & \textbf{Ovr} & \textbf{Rec} & \textbf{F1} & \textbf{Ord} & \textbf{Rel} & \textbf{Ovr} & \textbf{Rec} & \textbf{F1} & \textbf{Rel} & \textbf{Ovr} \\
    \midrule
    \multirow{6}[2]{*}{GPT-4o} & Single-Shot & 80.1  & \underline{69.1} & 90.8  & 74.8  & 39.7  & 38.0    & 25.6  & 77.9  & 55.6  & 90.5  & 90.5  & 94.0     & 82.9 \\
          & Rule-Based & 65.7  & 57.5  & 82.5  & 69.8  & 44.8  & 44.3  & 32.4  & \underline{92.2} & 61.1  & 59.7  & 59.7  & 71.1  & 69.5 \\
          & M2IO-Base-3B & 76.0     & 55.0     & 90.9  & 71.3  & 32.7  & 31.9  & 23.1  & 86.0     & 55.5  & 70.7  & 70.7  & 86.8  & 76.2 \\
          & M2IO-Base-72B & \underline{83.7} & \textbf{69.3} & \textbf{97.6} & \textbf{76.6} & \textbf{52.7} & \textbf{47.3} & \underline{38.6} & \textbf{93.9} & \textbf{63.3} & \underline{92.8} & \textbf{93.3} & \textbf{97.1} & \textbf{84.4} \\
          & M2IO-SFT-3B & 75.7  & 64.6  & \underline{98.0}     & 75.5  & 40.9  & 38.6  & 30.3  & 81.8  & 57.5  & 92.1  & 92.1  & 93.5  & 83.2 \\
          & M2IO-R1-3B & \textbf{84.2} & 68.4  & 97.4  & \underline{76.3} & \underline{52.6} & 46.5  & \textbf{39.4} & 90.5  & \underline{62.6} & \textbf{93.3} & \textbf{93.3} & \underline{97.0 } & \textbf{84.4} \\
    \midrule
    \multirow{6}[2]{*}{Qwen2.5-VL-72B} & Single-Shot & \underline{83.3} & \underline{65.6} & 90.8  & 73.7  & 35.2  & 28.6  & 25.6  & 90.9  & 56.4  & 94.3  & 94.4  & 89.3  & 83.0  \\
          & Rule-Based & 64.6  & 56.6  & 84.9  & 69.9  & \textbf{55.5} & 43.2  & 31.0     & \underline{92.2} & 60.7  & 51.3  & 58.0     & 67.5  & 68.4 \\
          & M2IO-Base-3B & 80.9  & 52.6  & 88.3  & 69.8  & 34.9  & 33.1  & 23.5  & 89.3 & 56.6  & 60.9  & 70.3  & 87.2  & 76.4 \\
          & M2IO-Base-72B & \textbf{83.7} & 63.4  & \underline{95.9} & \underline{74.4} & 49.4  & \textbf{45.6} & \underline{32.6} & \textbf{93.5}  & \textbf{61.7} & 96.7  & \underline{94.9} & \textbf{95.9} & \textbf{84.7} \\
          & M2IO-SFT-3B & 78.8  & 61.8  & 93.9  & 73.5  & 45.2  & 39.7  & 32.4  & 84.7  & 58.8  & \underline{95.0} & 94.7  & 90.5  & 83.3 \\
          & M2IO-R1-3B & \underline{83.3} & \textbf{66.0} & \textbf{96.7} & \textbf{75.2} & \underline{52.3} & 45.2  & \textbf{36.3} & 90.3  & \textbf{61.7} & \textbf{95.4} & \textbf{95.4} & \underline{93.0} & \underline{84.1} \\
    \midrule
    \multirow{6}[2]{*}{Qwen2.5-VL-7B} & Single-Shot & 62.2  & 40.2  & 90.4  & 65.4  & \underline{49.4} & 27.0     & 16.4  & \underline{92.0} & 53.2  & 86.3  & 86.7  & 89.0    & 79.6 \\
          & Rule-Based & 56.3  & 45.0    & 82.7  & 64.7  & 41.5  & 38.1  & 26.0    & 89.1  & 56.7  & 68.1  & 69.7  & 76.3  & 72.2 \\
          & M2IO-Base-3B & 75.2  & 49.7  & 90.2  & 67.7  & 36.6  & 33.2  & 23.6  & 91.3  & 55.7  & 58.8  & 68.8  & 86.8  & 74.6 \\
          & M2IO-Base-72B & \underline{77.2} & \textbf{66.9} & \underline{95.7} & \textbf{73.4} & \textbf{51.1} & \textbf{45.9} & \textbf{34.9} & \underline{92.0}    & \textbf{60.7} & 87.8  & 89.0    & 93.7  & 81.3 \\
          & M2IO-SFT-3B & 74.3  & \underline{59.2} & 94.5  & 71.2  & 39.8  & 34.8  & 27.4  & 85.8  & 55.7  & \underline{91.6} & \underline{90.9} & \underline{95.3} & \underline{82.2} \\
          & M2IO-R1-3B & \textbf{79.3} & 59.1  & \textbf{95.9} & \underline{71.5} & 48.7  & \underline{39.2} & \underline{28.3} & \textbf{92.1} & \underline{58.0} & \textbf{93.1} & \textbf{92.6} & \textbf{95.7} & \textbf{82.7} \\
    \bottomrule
    \end{tabular}%
  \label{tab:main_table}
\end{table*}%

\paragraph{Implementation Details}
We implement BGE-M3 \citep{bge-m3} as the text retriever for all baselines, which is used to retrieve relevant documents and their associated multi-modal images. 

For the text generation module in M2IO, we evaluate three models for answer generation: GPT-4o \citep{gpt4o}, Qwen2.5-VL-72B-Instruct, and Qwen2.5-VL-7B-Instruct. 

We adopt BGE-M3 and BGE-VL-base \citep{zhou2024megapairs} as the embedding models %for the bipartite graph matching algorithm.
for the rule-based strategy.

We adopt only open-source base models for our M2IO-Base image inserter to ensure reproducibility and fair comparison. 
Specifically, we employ two MLLM-based variants as the inserters for M2IO-Base: Qwen2.5-VL-3B-Instruct for M2IO-Base-3B and Qwen2.5-VL-72B-Instruct for M2IO-Base-72B.

In addition, the inserters of both M2IO-SFT-3B (implemented as Inserter-SFT-3B) and M2IO-R1-3B (implemented as Inserter-R1-3B) are initialized from Qwen2.5-VL-3B-Instruct and trained on the dataset discussed in Section~\ref{sub_data_se}.
Inserter-SFT-3B is trained via supervised fine-tuning (SFT) with the LLaMA-Factory framework \citep{zheng2024llamafactory}, and Inserter-R1-3B is trained within the EasyR1 framework \citep{sheng2024hybridflow, zheng2025easyr1} employing the GRPO algorithm. Further implementation details can be found in the Appendix.

\subsection{Main Results}
\label{sec: main_results}
\paragraph{Performance on MRAMG-Bench.} Table~\ref{tab:main_table} shows the evaluation results of various strategies, including the baselines and our M2IO series, on three MRAMG-Bench sub-datasets: Arxiv, Manual, and Web. The complete results on all six sub-datasets are provided in the Appendix. 

Our M2IO-R1-3B method consistently outperformed both Single-Shot and Rule-Based matching baselines across all three evaluated datasets. For instance, building upon GPT-4o generated text answers, M2IO-R1-3B demonstrated significant improvements on the Arxiv dataset compared to the Single-Shot method, increasing recall from 80.1 to 84.2, relevance from 90.8 to 97.4 and overall score from 74.8 to 76.3. Similarly, on the Manual dataset, it improves the order score from 25.6 to 39.4 and overall score from 55.6 to 62.6, over the Single-Shot approach. Furthermore, all four metrics on the Web dataset also show significant improvements compared to the Single-Shot and Rule-Based matching methods. 

These results not only underscore the effectiveness of our method but also demonstrate the superior performance of M2IO-R1-3B compared to traditional methods.

\paragraph{Improvement from Task Decomposition.} 
Compared to the Single-Shot method, our M2IO framework effectively exploits task decomposition, separating the generation of text answers and image insertion into two sequential sub-tasks.

Employing GPT-4o as the text answer generator, our M2IO-Base-72B, demonstrates a clear advantage over the Single-Shot GPT-4o baseline on the ArXiv dataset. Our method achieves superior results across three distinct metrics, most notably improving the relevance score from 90.8 for the baseline to 97.6. Surprisingly, when using Qwen2.5-VL-7B as the text answer generator, our smaller M2IO-Base-3B not only surpasses the Single-Shot baseline on the Arxiv dataset, but also achieves comparable performance on the Manual and Web datasets. 

These findings indicate that task decomposition effectively simplifies the image insertion challenge, thereby enhancing the overall quality of the generated multimodal output.

\paragraph{RL-based Reasoning Enhancement.} 
The effectiveness of RL fine-tuning is particularly evident when handling complex text answers, such as those from GPT-4o—that demand a deep understanding of logical flow. 

In this scenario, our base method, M2IO-Base-3B, initially underperforms the Single-Shot baseline. However, after RL training, the resulting Inserter-R1-3B not only surpasses the Single-Shot baseline but also achieves performance comparable to our state-of-the-art Qwen2.5-VL-72B inserter. This result highlights the remarkable parameter efficiency of our RL approach, enabling a compact 3B model to approach the advanced capabilities of a model over 20 times its size. 

Moreover, this underscores the superiority of our outcome-reward based RL method over SFT in cultivating the nuanced reasoning skills necessary for this task, as M2IO-R1-3B achieves significant improvements over M2IO-SFT-3B across nearly all metrics and datasets.

\begin{table}[htbp]
  \centering
  \caption{Comparison of inference latency and cost for three strategies on the MRAMG-Bench.}
    \begin{tabular}{l|cc}
    \toprule
    \multicolumn{1}{c|}{\textbf{Strategy}} & \textbf{Latency} (s) & \textbf{Cost} (\$) \\
    \midrule
    Single-Shot & 5.98  & 0.31 \\
    Rule-Based & 22.60  & \textbf{0.24} \\
    M2IO  & \textbf{4.34} & \textbf{0.24} \\
    \bottomrule
    \end{tabular}%
  \label{tab: efficiency}%
\end{table}%

\begin{table*}[t]
  \centering
  %\small
  \caption{Performance comparisons between the M2IO framework and baselines, where Inserter-SFT-3B and Inserter-R1-3B are trained on 80\% of the web-related datasets (\texttt{Wit}, \texttt{Web}, \texttt{Wiki}) and evaluated on the corresponding three MRAMG-Bench test sets. The best score is in \textbf{bold} and the second best is \underline{underlined}.}
    \begin{tabular}{l|cccc|ccccc|cccc}
    \toprule
    \multicolumn{1}{c|}{\multirow{2}[2]{*}{\textbf{Strategy}}} & \multicolumn{4}{c|}{\textbf{Arxiv}} & \multicolumn{5}{c|}{\textbf{Manual}}  & \multicolumn{4}{c}{\textbf{Web}} \\
          & \textbf{Rec} & \textbf{F1} & \textbf{Rel} & \textbf{Ovr} & \textbf{Rec} & \textbf{F1} & \textbf{Ord} & \textbf{Rel} & \textbf{Ovr} & \textbf{Rec} & \textbf{F1} & \textbf{Rel} & \textbf{Ovr} \\
    \midrule
    Single-Shot & 78.3  & 68.4  & 89.5  & 74.4  & 38.6  & 37.2  & 24.5  & 75.5  & 54.5  & 88.4  & 90.8  & \textbf{95} & 83.1 \\
    Rule-Based & 64.5  & 38.3  & 52.7  & 57.6  & 35.7  & 37.6  & 23.0  & 83.6  & 55.9  & 48.3  & 35.7  & 45.8  & 57.1 \\
    M2IO-Base-3B & 77.3  & 53.8  & 92.3  & 71.4  & 32.1  & 31.5  & 22.1  & 84.1  & 54.6  & 59.6  & 69.0  & 87.8  & 75.9 \\
    M2IO-Base-72B & \textbf{85.5} & \underline{68.5} & \underline{96.9} & \underline{76.2} & \textbf{50.9} & \textbf{45.6} & \textbf{35.4} & \textbf{90.9} & \textbf{61.4} & \underline{91.4} & \underline{92.4} & 94.4  & \underline{83.4} \\
    M2IO-SFT-3B & 78.0  & 57.4  & 88.8  & 71.4  & 36.6  & 27.9  & 15.7  & 61.3  & 48.0  & 88.4  & 88.9  & 90.2  & 81.5 \\
    M2IO-R1-3B & \underline{80.6} & \textbf{72.0} & \textbf{97.4} & \textbf{77.2} & \underline{42.3} & \underline{40.0} & \underline{29.1} & \underline{85.7} & \underline{58.0} & \textbf{94.0} & \textbf{94.0} & \underline{94.6} & \textbf{83.8} \\
    \bottomrule
    \end{tabular}%
  \label{tab: new_data}%
\end{table*}%

\subsection{Further Analysis}

\paragraph{Efficiency.} In addition to its state-of-the-art multimodal generation capabilities, we evaluated the practical efficiency of M2IO method. The analysis, presented in Table~\ref{tab: efficiency}, compares M2IO against both Single-Shot and Rule-Based baselines in terms of inference latency and monetary cost, where we disregard the computational overhead of our locally deployed 3B model and consider only the expenses from API calls as the cost metric. The results highlight M2IO's exceptional efficiency. 

Quantitatively, M2IO achieves a latency of just 4.34 seconds/instance, significantly outperforming both Single-Shot (5.98 seconds/instance) and Rule-Based (22.60 seconds/instance) approaches. Notably, M2IO is approximately 5.2 times faster than Rule-Based strategy. In terms of financial expenditure, M2IO is also highly economical, incurring a cost of 0.24 dollars/instance. This matches the most cost-effective baseline (Rule-Based) and represents a 22.6\% cost reduction compared to Single-Shot method. 

Our key finding is that M2IO does not force a trade-off between performance and efficiency; it excels at both. This remarkable efficiency, combined with its superior output quality, underscores the viability of M2IO for practical, real-world deployment.

\paragraph{Out-of-Domain Generalization.}
We evaluate the out-of-domain (OOD) generalization of our M2IO-R1 method on two benchmarks:
\begin{itemize}
    \item 
    \textbf{Web-Style-Only MRAMG-Bench.} 
    %Comprises the Web, Wiki, and Wit subsets for RL/SFT training.
    This setting comprises the \texttt{Web}, \texttt{Wiki}, and \texttt{Wit} subsets for RL/SFT training. 
    The models are trained on these subsets and evaluated on the full MRAMG-Bench.
    \item 
    \textbf{M2RAG Dataset.} An MRAMG task dataset derived from the ELI5 dataset \citep{fan2019eli5}. We train our models on the full MRAMG-Bench and test their generalization on M2RAG.
\end{itemize}

The results of our out-of-domain generalization experiments are detailed in Table~\ref{tab: new_data} and Table~\ref{tab: m2rag_ftii}. 

We observe two key sets of findings:

\begin{itemize}
    \item \textbf{Training on Web-Style Data Only:} First, when trained exclusively on web-style data (reducing the training set from 2.4k to 1.4k samples), the model's recall understandably decreases on non-web datasets. However, we note two significant advantages of our RL approach in this challenging, low-data setting: 
    \begin{itemize}
        \item[(1)] The F1 score on the Arxiv test set paradoxically improves, suggesting our RL method mitigates potential overfitting from the original training data.
        \item[(2)] The performance gap between the RL-tuned model and the SFT-only model widens, highlighting the superior generalization capability conferred by RL. For instance, on the Manual dataset, our M2IO-R1-3B achieves an F1 score of 40.0, markedly higher than M2IO-Base-3B (31.5) and M2IO-SFT-3B (27.9).
    \end{itemize}
    % (1) The F1 score on the Arxiv test set paradoxically improves, suggesting our RL method mitigates potential overfitting from the original training data. (2) The performance gap between the RL-tuned model and the SFT-only model widens, highlighting the superior generalization capability conferred by reinforcement learning. For instance, on the Manual dataset, our M2IO-R1-3B achieves an F1 score of 40.0, markedly higher than M2IO-Base-3B (31.5) and M2IO-SFT-3B (27.9).

    \item \textbf{Generalizing to the M2RAG Dataset:} When tested on the M2RAG dataset, M2IO-R1-3B outperforms the Rule-Based baseline. Furthermore, RL training significantly narrows the performance gap between M2IO-R1-3B and both the strong Single-Shot baseline and the much larger 72B parameter model, demonstrating its effectiveness in generalizing to new data distributions.
\end{itemize}

\paragraph{Image Insertion Enhancement.}
To evaluate how our GRPO-based RL method enhances the model's image insertion capabilities, we use the FTII-Bench dataset \citep{ruan2024ftii}, a benchmark for streaming image insertion derived from English news articles from the BBC.

Our results demonstrate a clear progression: while the M2IO-Base-3B strategy initially underperforms a Rule-Based Matching baseline, both SFT and RL deliver significant performance boosts. The improvement from RL is particularly notable. Consequently, our final M2IO-R1-3B strategy significantly outperforms the Rule-Based method. This enhancement also narrows the performance gap (in terms of F1 score) with the much larger 72B model from 35.9 to 12.2.

\paragraph{Ablation Study.} We conduct an ablation study on the hyperparameter $\alpha$ in the reward function $\mathcal{R}_{\text{answer}}$, which balances image recall and positional accuracy. 

We evaluate $\alpha \in \{0.0, 0.2, 0.4, 0.5, 0.6, 0.8, 1.0\}$ and find that $\alpha=0.8$ achieves the most optimal and balanced performance. Excluding the extreme cases of $\alpha=0$ and $\alpha=1$, our M2IO-R1-3B strategy consistently outperforms all baselines, demonstrating its robustness. 

For instance, with $\alpha=0.4$, M2IO-R1-3B strategy's F1 score on the Manual dataset (43.2) significantly surpasses M2IO-SFT (38.6).  
On the ArXiv dataset, M2IO-R1-3B with $\alpha = 0.5$ achieves an F1 score of 70.4, outperforming the larger M2IO-Base-72B method (69.3). These results
confirm the effectiveness of our method. Detailed experimental results can be found in the Appendix.

\begin{table}[htbp]
%\small
  \centering
  \caption{Performance comparisons between M2IO and the baselines on M2RAG and FTII-Bench.}
    \begin{tabular}{l|cc|cc}
    \toprule
    \multicolumn{1}{c|}{\multirow{2}[2]{*}{\textbf{Strategy}}} & \multicolumn{2}{c|}{\textbf{M2RAG}} & \multicolumn{2}{c}{\textbf{FTII-Bench}} \\
          & \textbf{Pos} & \textbf{Rel} & \textbf{F1} & \textbf{Pos} \\
    \midrule
    Single-Shot & \underline{80.1} & 77.1  &  --     & --  \\
    Rule-Based & 55.7  & 54.4  & 29.8  & 28.8  \\
    M2IO-Base-3B & 72.0  & 72.6  & 29.0  & 14.4  \\
    M2IO-Base-72B & \textbf{82.3} & \textbf{83.5} & \textbf{64.9} & \textbf{35.2} \\
    M2IO-SFT-3B & 67.4  & 76.8  & 46.3  & 28.8  \\
    M2IO-R1-3B & 70.5  & \underline{79.4} & \underline{52.7} & \underline{31.8} \\
    \bottomrule
    \end{tabular}%
  \label{tab: m2rag_ftii}%
\end{table}%

%% file: Conclusion.tex
\section{Conclusion}
% In this paper, we introduced M2IO-R1, a novel framework integrating MRAMG task with RL.
% By decomposing the multimodal response pipeline into four stages—retrieval, text generation, image insertion, and merging—M2IO-R1 produces controllable and interpretable multimodal outputs.
% Central to the framework is an RL–based inserter trained with GRPO, which enables precise and semantically aligned visual integration.
% Extensive experiments across multiple benchmarks demonstrate that M2IO-R1 consistently outperforms competitive baselines. 
% Notably, the lightweight M2IO-R1-3B delivers a “punching above its weight” effect, matching or even surpassing the 72B-parameter model while incurring only a fraction of the computational cost and reducing inference latency.

In this paper, we introduced M2IO-R1, a novel framework that integrates the MRAMG task with RL strategy. By decomposing the pipeline into four distinct stages—retrieval, text answer generation, image insertion, and merging—M2IO-R1 produces controllable and interpretable multimodal outputs. 

A key contribution of our work is the RL-based inserter. Trained with the outcome-reward-based GRPO, this component achieves precise and semantically coherent visual integration, ensuring that images are placed in the most relevant locations within the text. Our extensive experiments across multiple benchmarks demonstrate that M2IO-R1 consistently outperforms competitive baselines. 

Notably, the lightweight M2IO-R1-3B strategy delivers a ``punching above its weight'' effect. It matches or even surpasses the performance of models with orders of magnitude more parameters, such as the 72B-parameter baseline (M2IO-Base-72B), while incurring only a fraction of the computational cost and significantly reducing inference latency. This result highlights the efficiency and effectiveness of our approach, demonstrating that carefully designed frameworks can achieve state-of-the-art performance without relying on massive model scale.

%% file: Appendix.tex
\section{Appendix}

\subsection{Dataset Details}
In this section, we provide a detailed introduction to the datasets used in this work.

\paragraph{MRAMG-Bench \citep{yu2025mramg}} is a meticulously curated, human‑annotated benchmark for the Multimodal Retrieval‑Augmented Multimodal Generation (MRAMG) task. It provides ground‑truth annotations for both textual answers and image placements, comprising 4,800 QA pairs, 4,346 documents, and 14,190 images. The benchmark integrates six sub‑datasets—MRAMG‑Wit, MRAMG‑Wiki, MRAMG‑Web, MRAMG‑Arxiv, MRAMG‑Recipe, and MRAMG‑Manual—spanning three domains: Web, Academia, and Lifestyle. By incorporating hierarchical difficulty levels and multi‑image reasoning scenarios, MRAMG-Bench offers a robust foundation for evaluating the accuracy, coherence, and grounding capabilities of multimodal generation systems.

\paragraph{FTII-Bench \citep{ruan2024ftii}} is a bilingual benchmark designed for the novel Flow Text with Image Insertion (FTII) task. It is constructed from 625 high-quality, image-text news articles (318 Chinese and 307 English), collectively forming 10,231 questions across two distinct formats: single-choice and flow-insertion. The benchmark spans 10 different news domains and introduces hierarchical difficulty levels by carefully controlling the semantic relevance of distractor images. FTII-Bench provides a robust framework for comprehensively evaluating a model's synergistic capabilities in long-text interpretation, multi-image comprehension, and complex instruction following, challenging even the most advanced systems.
We select the English subset of the flow-insertion task, filtering out samples containing more than six images. To ensure consistency with the MRAMG test sets, we retain the same ground-truth images while maintaining a 1:1 ratio between distractor and ground-truth images.
After filtering, the dataset sizes are as follows: FTII-1 (Level 1) contains 461 samples, FTII-2 (Level 2) contains 423 samples, and FTII-3 (Level 3) contains 201 samples.

\paragraph{M2RAG \citep{ma2024multi}} is a comprehensive benchmark meticulously designed for the MRAMG task. %It evaluates a model's ability to process multimodal  content and generate coherent, mixed-layout responses of text and images. 
It evaluates the model’s capability to process multimodal content and to generate coherent, interleaved responses that seamlessly integrate text and images.
The dataset is derived from queries in the ELI5 dataset \citep{fan2019eli5}. For each query, relevant web documents were crawled and automatically filtered, yielding 750 queries paired with corresponding reference texts and auxiliary images. The benchmark spans 10 distinct topics, including Science, Health, and Politics, ensuring diverse and real-world scenarios. Due to the lack of ground truth answers, the original dataset inherently adopts an LLM-as-a-Judge evaluation paradigm, focusing on multimodal factors such as image–text semantic relevance and consistency and image position accuracy.
%该数据集对于每个query并没有提供参考答案，所以采用了llm as judge的评估指标，例如衡量图文一致性和图片位置准确性M2RAG offers a foundation model-based evaluation framework, which employs a suite of fine-grained text and multi-modal metrics (e.g., Image Relevance) to provide a robust foundation for analyzing the retrieval, reasoning, and multi-modal synthesis capabilities of advanced foundation models.

\subsection{Metric Details}

This section details the evaluation metrics used in our study.

    \paragraph{Recall (Rec)}  measures the proportion of correct images in a multimodal answer relative to the total number of images in the ground truth, thereby evaluating the effectiveness of the answer in incorporating relevant and informative visual content.
    %measures the percentage of correct images in the multimodal answer relative to the total number of images in the ground truth, evaluating whether the answer effectively includes useful image information. 
    It is computed as:
    \begin{equation}
        \text{Recall} = \frac{\text{True Positives}}{\text{True Positives} + \text{False Negatives}},
    \end{equation}

    where False Negatives are the images in the ground truth that were omitted in the generated multimodal answer.

     \paragraph{F1-Score (F1)} serves as the harmonic mean of Precision and Recall, offering a balanced evaluation of image quality in multimodal answers. 
It is defined as:
\begin{equation}
    \text{F1-Score} = 2 \times \frac{\text{Precision} \times \text{Recall}}{\text{ Precision} + \text{Recall}} ,
\end{equation}
where
\begin{equation}
    \text{Precision} = \frac{\text{True Positives}}{\text{True Positives} + \text{False Positives}},
\end{equation}
with True Positives denoting correctly inserted images and False Positives referring to irrelevant images that were included.

    \paragraph{Order Score (Ord) \citep{yu2025mramg}} evaluates the alignment between the image sequence in the generated answer and the ground-truth sequence. We compute the score using the weighted edit distance between these two sequences. This metric is applied exclusively to the MRAMG-Recipe and MRAMG-Manual datasets, as their instances often contain multiple images where the sequential order is critical. The score is defined as follows:
    \begin{description}
        \item[\textbf{Ground-truth}] $\mathcal{I}^{*} = i_1^{*} \rightarrow i_2^{*} \rightarrow \cdots \rightarrow i_n^{*}$, where $i_j^{*}$ represents the image at the $j$-th position in the order.
        \item[\textbf{Answer}] $\mathcal{I} = i_1 \rightarrow i_2 \rightarrow \cdots \rightarrow i_m$, where $i_j$ is not necessarily in $\mathcal{I}^{*}$, and the number of inserted images $m$ is not necessarily equal to $n$.
    \end{description}
    Then, the scoring formula is:
    \begin{equation}
         \text{Score} = \frac{|\mathcal{I}^{*} \cap \mathcal{I}|}{n} \times \left( 1 - \frac{1}{p} \times \min\left(\frac{\text{dist}(\mathcal{I}^{*}, \mathcal{I})}{n}, p \right)\right).
    \end{equation}
    Here, $\text{dist}(\mathcal{S}, \mathcal{S}')$ represents the weighted edit distance between the sequences $\mathcal{S}$ and $\mathcal{S}'$, defined as the minimum total cost to transform $\mathcal{S}'$ into $\mathcal{S}$ using the following operations:
\begin{description}
    \item[\textbf{Insertion}] Insert an image from $\mathcal{S}$ that is missing in $\mathcal{S}'$. The operation cost is $p_1$.
    \item[\textbf{Deletion}] Delete an image from $\mathcal{S}'$ that is not present in $\mathcal{S}$. The operation cost is $p_2$.
    \item[\textbf{Substitution}] Replace an image in $\mathcal{S}'$ with the correct image from $\mathcal{S}$ at the corresponding position. The operation cost is $p_3$.
\end{description}
The operation costs typically satisfy $p_1 > p_2 > p_3$, and $p \geq p_1$ ensures the final score falls within the range $[0, 1]$. This weighted edit distance can be computed using dynamic programming with a time complexity of $O(mn)$.
    % Here, $\text{dist}(\mathcal{S}, \mathcal{S}')$ represents the weighted edit distance between string $\mathcal{S}$ and string $\mathcal{S}'$, i.e., the minimum total cost to transform string $\mathcal{S}'$ into string $\mathcal{S}$ through the following three operations:
    %     \begin{itemize}
    %         \item \textbf{String Insertion}: If $\mathcal{S}'$ is missing certain images, insert an image from $\mathcal{S}$ into a specific position in $\mathcal{S}'$. The operation cost is $p_1$.
    %         \item \textbf{String Deletion}: If $\mathcal{S}'$ contains extra irrelevant images, delete them. The operation cost is $p_2$.
    %         \item \textbf{String Substitution}: If the positions of images in $\mathcal{S}'$ do not match $\mathcal{S}$, substitute the image in $\mathcal{S}'$ with the corresponding image from $\mathcal{S}$. The operation cost is $p_3$.
    %     \end{itemize}
    % The weights generally satisfy $p_1 > p_2 > p_3$, and $p \geq p_1$ ensures the final score falls within the range $[0, 1]$. Weighted edit distance can be computed using dynamic programming, with a time complexity of $O(mn)$.
    %\end{itemize}
    
    \paragraph{Relevance (Rel)\citep{zhu2024murar,ma2024multi,yu2025mramg}}  evaluates the relevance of the inserted image to the query-answer pair, specifically assessing whether the content described by the image is meaningfully related to the content of the QA. 
    This metric adopts an LLM‑as‑a‑Judge paradigm, in which a large language model evaluates each image in the answer by assigning a score on a 1–5 scale, thereby providing an automated and consistent measure of quality.
    %This metric assigns a score to each image appearing in the answer, with scores ranging from 1 to 5. we adopt an LLM‑based evaluation approach, %Specifically, we employ \textbf{GPT-4o-mini} to evaluate the \textbf{Rel} metric.
    
    % \textbf{Position Score (Pos)} is used to assess the appropriateness of the image placement in the multimodal answer. Specifically, FTII-Bench provide ground truth answers for queries. We evaluate the Pos metric on FTII-Bench for each query based on corresponding ground truth answers. Let $\mathcal{I}=\{i_1,i_2,\dots,i_m\}$ denote the image insertion result, and let $\mathcal{I}^*=\{i_1^*,i_2^*,\dots,i_m^*\}$ denote the ground-truth image sequence. Then, we compute the Pos metric as:
    % \begin{equation}
    %     \textbf{Pos} = \begin{cases}
    %         0, &\text{if $\mathcal{I}$ is empty but $\mathcal{I}^*$ is not empty;}
    %         \\
    %         1, &\text{if $\mathcal{I}$ is empty and $\mathcal{I}^*$ is empty;}
    %         \\
    %         \frac{\sum_{i=1}^{m} P_i}{m}, &\text{otherwise.}
    %     \end{cases}
    % \end{equation}
    % For each image, we compute image position score $P_i$ as:
    % \begin{equation}
    %     \textbf{$P_i$} = \begin{cases}
    %         1, &\text{if $\mathcal{I}_i = \mathcal{I}_i^*$;}
    %         \\
    %         0.5, &\text{if $\mathcal{I}_i \neq \mathcal{I}_i^*$ and $\mathcal{I}_i \in \mathcal{I}^*$;}
    %         \\
    %         0, &\text{otherwise.}
    %     \end{cases}
    % \end{equation}
    % When evaluating the Pos metric on M2RAG datasets, we leverage GPT-4o-mini as the evaluator with meticulous designed prompt (See \ref{Position}).
    \paragraph{Position Score (Pos)} evaluates the appropriateness of image placement in the generated answer. It measures whether the correct images are placed after the correct sentences.

To define the score, let the generated sequence of images be $\mathcal{I}=\{i_1, i_2, \dots, i_m\}$ and the ground-truth sequence be $\mathcal{I}^*=\{i_1^*, i_2^*, \dots, i_m^*\}$. Here, $m$ is the total number of sentences, and each element $i_j$ (or $i_j^*$) represents the image inserted after the $j$-th sentence. If no image is inserted, the element is empty, i.e., $i_j = \emptyset$.

The overall Pos score is calculated based on a per-position score, $P_j$, for each of the $m$ positions. The score $P_j$ for a single position is defined as:
\begin{equation}
    P_j = \begin{cases}
        1,   & \text{if } i_j = i_j^*; \\
        0.5, & \text{if } i_j \neq i_j^* \text{ and } i_j \in \mathcal{I}^*; \\
        0,   & \text{otherwise.}
    \end{cases}
\end{equation}
This rewards placing the correct image in the correct slot ($P_j=1$), gives partial credit for placing a correct image in the wrong slot ($P_j=0.5$), and gives no credit for using an incorrect image or omitting a required one.

The final Pos metric averages these per-position scores, with special handling for cases where no images are generated:
\begin{equation}
        \text{Pos} = \begin{cases}
        1,   & \text{if both } \mathcal{I} \text{ and } \mathcal{I}^* \text{ are empty;} \\
        0,   & \text{if } \mathcal{I} \text{ is empty but } \mathcal{I}^* \text{ is not;} \\
        \frac{1}{m} \sum_{j=1}^{m} P_j, & \text{otherwise.}
    \end{cases}
\end{equation}

We compute this metric on the FTII-Bench dataset using its provided ground-truth answers. %For datasets like \textbf{M2RAG} that lack ground-truth positions, we employ \textbf{GPT-4o-mini} as an automated evaluator, guided by a meticulously designed prompt.
%For datasets such as M2RAG, which lack ground‑truth positional information, we employ an LLM as an automated evaluator, guided by a carefully designed prompt.
For datasets such as M2RAG, where ground‑truth positional information is unavailable, we adopt an LLM‑as‑a‑Judge evaluation approach, employing a carefully crafted prompt to ensure reliable and consistent assessment. This metric serves as a comprehensive measure of overall quality.

    %\yqh{介绍不同，以及放上prompt}
    
    \paragraph{Overall Score (Ovr)} assesses the overall quality and helpfulness of a multimodal answer compared to the ground truth. In addition to evaluating multimodal quality, we incorporate BERTScore \citep{lin-2004-rouge} and ROUGE‑L \citep{zhang2019bertscore} as well‑established metrics to assess the textual quality of the generated answers.
    The final Ovr is calculated as the mean of F1, Rel, BERTScore, and ROUGE‑L, with Ord additionally incorporated when order information is available
    %Importantly, we leverage \textbf{Bert score} and \textbf{RougeL score} as metrics to evaluate the quality of generated answer text of the text generators (GPT-4o, Qwen2.5-VL-72B, Qwen2.5-VL-7B). \textbf{Ovr} is computed as the average of \textbf{F1}, \textbf{Ord}, \textbf{Rel}, \textbf{Bert score} and \textbf{RougeL score}.
to the ground truth.

\subsection{Training Details}

In this section, we provide the full training details of our inserters. 

The inserters of M2IO‑SFT (i.e., Inserter‑SFT‑3B) and M2IO‑R1 (i.e., Inserter‑R1‑3B) are initialized from Qwen2.5‑VL‑3B‑Instruct and trained on the M2IO-Inserter dataset described in Section~\ref{sub_data_se}.

Inserter-SFT-3B is trained via supervised fine-tuning (SFT) using the LLaMA-Factory framework \citep{zheng2024llamafactory}, with full-parameter fine-tuning. The training is conducted for 3 epochs with a learning rate of 1e-6 and an effective global batch size of 64.

Inserter-R1-3B is trained within the EasyR1 framework \citep{sheng2024hybridflow, zheng2025easyr1} using the Group Relative Policy Optimization (GRPO) algorithm \citep{shao2024deepseekmath}. The training is conducted with a global batch size of 32, a learning rate of 1e-6, and a KL penalty coefficient set to $0.01$.

\subsection{Prompt Details}

The LLM prompts used in this study are presented below.

\begin{tcolorbox}[title=Evaluation Prompt on Relevance]
\begin{lstlisting}[numbers=none]
# Input
Query: {query_str}
Answer: {answer_str}
Image Context: {context_str_list}
Image Caption: {caption_str_list}
Image number to be rated: {image_number}

# Task
Imagine you are a multimodal QA evaluation expert. Your task is to evaluate the relevance the selected images within an answer to the given query and the overall quality of the answer. Specifically, the answer contains both text and images. You need to assess whether the selected images are relevant to the QApair in terms of content.

## Answer Input Format
[text_context_1] <img_1> [text_context_2] <img_2>...

Explanation:
Each ``[text_context_x]'' is a piece of pure text context, and each <img> represents an image. The images will be provided in the same order as the placeholders <img>.

## Image Context Input Format
[context_above] <img> [context_bottom]
Explanation: The image to be evaluated is provided along with its preceding and following context in placeholder form.

# Scoring Criteria of Relevance
When scoring, strictly adhere to the following standards, with a range of 1 to 5:
- 1 point: Completely unrelated: The images in the answer have no connection to the main content of the query and answer, and are irrelevant overall.
- 2 points: Weakly related: The images in the answer have a very tenuous connection to the main content of the query and answer.
- 3 points: Partially related: The images in the answer are somewhat connected to part of the content of the query and answer.
- 4 points: Mostly related: The images in the answer have a fairly clear connection to the main content of the query and answer.
- 5 points: Highly related: The images in the answer are highly relevant to the content of the query and answer.

Provide a brief reason for the evaluation along with a score from 1 to 5. Ensure you do not use any evaluation criteria beyond the query and answer.

# Output Format
Please output two lines for each measure: the first line is your reasoning for the score, and the second line is the score. Strictly follow this format without any additional content.

# requirements
1. There must be an integer output for each score (1-5).
2. There must be a <relevance_score> tag for the relevance score and a <effective_score> tag for the effectiveness score, and a <overall_quality_score> tag for the overall quality score.

# Output Example
Highly relevant: The images in the answer show the number of pillars in front of the gate, which perfectly match the query about the number of pillars. All images in the answer are highly relevant to the content of the query and answer.
<relevance_score>5</relevance_score>
\end{lstlisting}
\end{tcolorbox}

\begin{tcolorbox}[title=Evaluation Prompt on Position Score ]
\label{Position}
\begin{lstlisting}[numbers=none]

# Input
Query: {query_str}
Answer: {answer_str}
Image Context: {context_str_list}
Image Caption: {caption_str_list}
Image number to be rated: {image_number}

# Task
Imagine you are a multimodal problem-solving expert tasked with evaluating whether the position of each selected image within an answer to the given query is appropriate.
## Requirements
1. If there are repeated images in the answer, only evaluate the first occurrence, and for repeated images, rate them as 0 (except for the first occurrence).
## Answer Input Format
[text_context_1] <img_1> [text_context_2] <img_2>...
Explanation:
Each ``[text_context_x]'' is a segment of pure text context, and each <img> represents an image. The images will be presented in the same order as the placeholders <img>. 
## Image Context Input Format
[context_above] <img> [context_bottom] 
Explanation: The image under evaluation is provided along with its preceding and following contexts in placeholder form, corresponding to <img>.
# Revised Evaluation Criteria:
Strictly follow the criteria below to assign a score of 0 or 1:
- 0 points, Inappropriate Position: The image is irrelevant to both the preceding and following context, or the position of the image does not enhance content understanding or visual appeal. The insertion of the image does not align with the logical progression of the text and fails to improve the reading experience or information transmission.
- 1 point, Appropriate Position: The image is contextually relevant to at least one of the surrounding contexts (preceding or following), and it enhances content understanding or visual effect. The position of the image aligns with the logical flow of the text and is inserted appropriately, improving the overall information delivery. If the description of the image is detailed, it further clarifies the connection between the image and the text, enhancing the overall expressive effect.
# Output Format
Provide a brief justification for the evaluation and a score of either 0 or 1 for each unique image. Ensure no evaluation criteria beyond the provided query and answer are used.
For images that appear multiple times, evaluate only the first occurrence and do not provide additional scores for repeated occurrences.
Please output two lines for each image: the first line is your reasoning for the score, and the second line is the score. Strictly follow this format without any additional content.

# Output Example
<img_1> shows a decorative lamp post, but the surrounding context describes architectural features unrelated to the lamp post. The image placement disrupts the logical flow and fails to enhance understanding.
<img_1_score>0</img_1_score>
<img_2> depicts three pillars in front of a gate, and its context discusses the number of pillars in front of the gate. Therefore, the position is appropriate.  
<img_2_score>1</img_2_score>

\end{lstlisting}
\end{tcolorbox}
\begin{tcolorbox}[title= Training Prompt for M2IO‑R1]
\begin{lstlisting}[basicstyle=\ttfamily\normalsize, breaklines=true, numbers=none]
You are an expert in inserting images into texts, specializing in analyzing the relationship between images and texts and optimizing their alignment for enhanced readability and information clarity.
 ## Task
Given a question, a dictionary of ground truth answers (where keys are sentence indices and values are sentence content), and a list of candidate images (with captions or contexts), your task is to:
Assess image relevance and select suitable ones. Determine optimal image placements. Output only the selected image-sentence dict, where key is the image ID and value is the sentence index after which the image should be placed.
## Format
 Within <think> tags, explain your reasoning, covering: Image content analysis. Relevance to the text. Image placement and justification. Ensuring seamless multimodal integration.
In the <answer> tags, return a dictionary where each key is a selected image ID (e.g., ``image20'') and the value can only be a single sentence index (integer) indicating the exact sentence after which the image should be inserted.If no image is selected, return an empty dictionary.
Note: Only output the selected images and their respective placements.
Question: {question}
Ground truth answer dict: {ground_truth_dict}
Candidate images information: {imgs_info}
\end{lstlisting}
\end{tcolorbox}

\begin{tcolorbox}[title= Text Answer Generation Prompt for M2IO‑R1]
\begin{lstlisting}[basicstyle=\ttfamily\normalsize, breaklines=true, numbers=none]
# Input:
Question:{query_str}
Context:{context_str}

# Task
Imagine you are a text QA expert. You will be provided with a plain text context and a query related to that context. Your task is to answer the query based solely on the content of the context. Ensure that your answer does not include any additional information outside the context . Please note that your answer should be in pure text format.

# Output Format
Provide the answer in pure text format. Do not include any information beyond what is contained in the context.
\end{lstlisting}
\end{tcolorbox}

\begin{tcolorbox}[title= Image Insertion Prompt for M2IO‑R1]
\begin{lstlisting}[basicstyle=\ttfamily\normalsize, breaklines=true, numbers=none]
You are an expert in inserting images into texts, specializing in analyzing the relationship between images and texts and optimizing their alignment for enhanced readability and information clarity.
## Task
Given a question, a dictionary of ground truth answers (where keys are sentence indices and values are sentence content), and a list of candidate images (with captions or contexts), your task is to:
Assess image relevance and select suitable ones. Determine optimal image placements. Output only the selected image-sentence dict, where key is the image ID and value is the sentence index after which the image should be placed.
## Format
Within <think> tags, explain your reasoning, covering: Image content analysis. Relevance to the text. Image placement and justification. Ensuring seamless multimodal integration.
In the <answer> tags, return a dictionary where each key is a selected image ID (e.g., ``image20'') and the value can only be a single sentence index (integer) indicating the exact sentence after which the image should be inserted.If no image is selected, return an empty dictionary.
Note: Only output the selected images and their respective placements.
Question: {question}
Ground truth answer dict: {ground_truth_dict}
Candidate images information: {imgs_info}
\end{lstlisting}
\end{tcolorbox}

\begin{tcolorbox}[title= Image Insertion Prompt for M2IO‑Base and M2IO-SFT]
\begin{lstlisting}[basicstyle=\ttfamily\normalsize, breaklines=true, numbers=none]
You are an expert in inserting images into texts, specializing in analyzing the relationship between images and texts and optimizing their alignment for enhanced readability and information clarity.
## Task
Given a question, a dictionary of ground truth answers (where keys are sentence indices and values are sentence content), and a list of candidate images (with captions or contexts), your task is to:
Assess image relevance and select suitable ones. Determine optimal image placements. Output only the selected image-sentence dict, where key is the image ID and value is the sentence index after which the image should be placed.
## Format
Return a dictionary where each key is a selected image ID (e.g., ``image20'') and the value can only be a single sentence index (integer) indicating the exact sentence after which the image should be inserted. If no image is selected, return an empty dictionary.
Note: Only output the final answer in a single dict, with the selected image ids (e.g., ``image20'') as keys and their respective placements sentence index as value (e.g., ``2'').
Question: {question}
Ground truth answer dict: {ground_truth_dict}
Candidate images information: {imgs_info}
\end{lstlisting}
\end{tcolorbox}

%\newpage
\subsection{Additional Experiment Results}
In this section, we first present the complete experimental results across all benchmark datasets to provide a comprehensive view of the M2IO framework’s performance compared to baselines. We then conduct an ablation study to specifically evaluate the impact of the hyperparameter $\alpha$ within the reward function $\mathcal{R}_{\text{answer}}$ during the training of the Inserter-R1-3B model.
% Table generated by Excel2LaTeX from sheet 'res_paper_full_v2'
\begin{table*}[h]
  \centering
  \setlength{\tabcolsep}{3pt} % 您可以取消注释以调整列间距
  \caption{
  Performance comparisons between the M2IO framework and baselines on the three MRAMG-Bench datasets (\texttt{Arxiv}, \texttt{Manual}, \texttt{Recipe}).
  %Performance comparisons between the M2IO framework and baselines on the Arxiv, Manual, and Recipe datasets.
  }
    \begin{tabular}{c|l|cccc|ccccc|ccccc}
    \toprule
    \multirow{2}[2]{*}{\textbf{T‑generator}} & \multicolumn{1}{c|}{\multirow{2}[2]{*}{\textbf{Strategy}}} & \multicolumn{4}{c|}{\textbf{Arxiv}} & \multicolumn{5}{c|}{\textbf{Manual}} & \multicolumn{5}{c}{\textbf{Recipe}} \\
          &       & \textbf{Rec} & \textbf{F1} & \textbf{Rel} & \textbf{Ovr} & \textbf{Rec} & \textbf{F1} & \textbf{Ord} & \textbf{Rel} & \textbf{Ovr} & \textbf{Rec} & \textbf{F1} & \textbf{Ord} & \textbf{Rel} & \textbf{Ovr} \\
    \midrule
    \multirow{6}[2]{*}{GPT-4o} & Single-Shot & 80.1  & 69.1  & 90.8  & 74.8  & 39.7  & 38.0  & 25.6  & 77.9  & 55.6  & 50.1  & 45.9  & 33.0  & 77.2  & 59.7 \\
          & Rule-Based & 65.7  & 57.5  & 82.5  & 69.8  & 44.8  & 44.3  & 32.4  & 92.2  & 61.1  & 67.7  & 52.0  & 39.0  & 88.6  & 64.3 \\
          & M2IO-Base-3B & 76.0  & 55.0  & 90.9  & 71.3  & 32.7  & 31.9  & 23.1  & 86.0  & 55.5  & 57.8  & 47.4  & 37.8  & 86.2  & 62.7 \\
          & M2IO-Base-72B & 83.7  & 69.3  & 97.6  & 76.6  & 52.7  & 47.3  & 38.6  & 93.9  & 63.3  & 67.9  & 55.3  & 42.2  & 88.9  & 65.7 \\
          & M2IO-SFT-3B & 75.7  & 64.6  & 98.0  & 75.5  & 40.9  & 38.6  & 30.3  & 81.8  & 57.5  & 62.7  & 57.1  & 45.1  & 91.2  & 67.1 \\
          & M2IO-R1-3B & 84.2  & 68.4  & 97.4  & 76.3  & 52.6  & 46.5  & 39.4  & 90.5  & 62.6  & 68.2  & 56.4  & 44.3  & 88.4  & 66.2 \\
    \midrule
    \multirow{6}[2]{*}{Qwen2.5-VL-72B} & Single-Shot & 83.3  & 65.6  & 90.8  & 73.7  & 35.2  & 28.6  & 25.6  & 90.9  & 56.4  & 52.5  & 40.8  & 24.3  & 86.1  & 57.6 \\
          & Rule-Based & 64.6  & 56.6  & 84.9  & 69.9  & 55.5  & 43.2  & 31.0  & 92.2  & 60.7  & 66.1  & 48.0  & 34.3  & 88.1  & 61.5 \\
          & M2IO-Base-3B & 80.9  & 52.6  & 88.3  & 69.8  & 34.9  & 33.1  & 23.5  & 89.3  & 56.6  & 58.3  & 46.0  & 36.2  & 88.3  & 61.5 \\
          & M2IO-Base-72B & 83.7  & 63.4  & 95.9  & 74.4  & 49.4  & 45.6  & 32.6  & 93.5  & 61.7  & 66.7  & 50.6  & 35.0  & 89.8  & 62.5 \\
          & M2IO-SFT-3B & 78.8  & 61.8  & 93.9  & 73.5  & 45.2  & 39.7  & 32.4  & 84.7  & 58.8  & 61.4  & 53.2  & 32.4  & 88.6  & 62.2 \\
          & M2IO-R1-3B & 83.3  & 66.0  & 96.7  & 75.2  & 52.3  & 45.2  & 36.3  & 90.3  & 61.7  & 67.8  & 53.2  & 39.8  & 85.4  & 63.1 \\
    \midrule
    \multirow{6}[2]{*}{Qwen2.5-VL-7B} & Single-Shot & 62.2  & 40.2  & 90.4  & 65.4  & 49.4  & 27.0  & 16.4  & 92.0  & 53.2  & 61.3  & 31.8  & 16.7  & 88.0  & 53.4 \\
          & Rule-Based & 56.3  & 45.0  & 82.7  & 64.7  & 41.5  & 38.1  & 26.0  & 89.1  & 56.7  & 46.8  & 41.2  & 27.1  & 85.6  & 56.9 \\
          & M2IO-Base-3B & 75.2  & 49.7  & 90.2  & 67.7  & 36.6  & 33.2  & 23.6  & 91.3  & 55.7  & 61.6  & 45.5  & 33.2  & 87.7  & 59.4 \\
          & M2IO-Base-72B & 77.2  & 66.9  & 95.7  & 73.4  & 51.1  & 45.9  & 34.9  & 92.0  & 60.7  & 57.0  & 44.6  & 34.1  & 86.3  & 59.1 \\
          & M2IO-SFT-3B & 74.3  & 59.2  & 94.5  & 71.5  & 39.8  & 34.8  & 27.4  & 85.8  & 55.7  & 55.1  & 45.0  & 31.7  & 92.0  & 59.9 \\
          & M2IO-R1-3B & 79.3  & 59.1  & 95.9  & 71.2  & 48.7  & 39.2  & 28.3  & 92.1  & 58.0  & 65.1  & 45.5  & 32.9  & 88.9  & 59.6 \\
    \bottomrule
    \end{tabular}%
   \label{tab:complete_old_data_1}
\end{table*}%

\begin{table*}[h]
  \centering
  \setlength{\tabcolsep}{5pt} % 您可以取消注释以调整列间距
  \caption{Performance comparisons between the M2IO framework and baselines on the three MRAMG-Bench datasets (\texttt{Web}, \texttt{Wiki}, \texttt{Wit}).}
    \begin{tabular}{c|l|cccc|cccc|cccc}
    \toprule
    \multirow{2}[2]{*}{\textbf{T‑generator}} & \multicolumn{1}{c|}{\multirow{2}[2]{*}{\textbf{Strategy}}} & \multicolumn{4}{c|}{\textbf{Web}} & \multicolumn{4}{c|}{\textbf{Wiki}} & \multicolumn{4}{c}{\textbf{Wit}} \\
          &       & \textbf{Rec} & \textbf{F1} & \textbf{Rel} & \textbf{Ovr} & \textbf{Rec} & \textbf{F1} & \textbf{Rel} & \textbf{Ovr} & \textbf{Rec} & \textbf{F1} & \textbf{Rel} & \textbf{Ovr} \\
    \midrule
    \multirow{6}[2]{*}{GPT-4o} & Single-Shot & 90.5  & 90.5  & 94.0  & 82.9  & 72.9  & 72.8  & 63.9  & 71.6  & 85.1  & 84.7  & 75.2  & 79.1  \\
          & Rule-Based & 59.7  & 59.7  & 71.1  & 69.5  & 53.0  & 52.9  & 51.4  & 63.5  & 61.1  & 60.8  & 59.9  & 61.3  \\
          & M2IO-Base-3B & 70.7  & 70.7  & 86.8  & 76.2  & 82.9  & 82.7  & 93.6  & 81.5  & 85.4  & 84.9  & 91.1  & 85.7  \\
          & M2IO-Base-72B & 92.8  & 93.3  & 97.1  & 84.4  & 96.4  & 95.3  & 93.1  & 84.5  & 96.4  & 95.0  & 93.8  & 92.4  \\
          & M2IO-SFT-3B & 92.1  & 92.1  & 93.5  & 83.2  & 98.4  & 97.7  & 94.7  & 85.5  & 98.3  & 94.9  & 93.3  & 93.0  \\
          & M2IO-R1-3B & 93.3  & 93.3  & 97.0  & 84.4  & 98.4  & 97.5  & 94.9  & 85.5  & 99.0  & 97.3  & 96.4  & 94.5  \\
    \midrule
    \multirow{6}[2]{*}{Qwen2.5-VL-72B} & Single-Shot & 94.3  & 94.4  & 89.3  & 83.0  & 98.8  & 95.2  & 94.1  & 84.0  & 99.0  & 96.7  & 92.3  & 93.0  \\
          & Rule-Based & 51.3  & 58.0  & 67.5  & 68.4  & 52.2  & 52.1  & 51.5  & 62.6  & 58.5  & 17.6  & 58.0  & 49.2  \\
          & M2IO-Base-3B & 60.9  & 70.3  & 87.2  & 76.4  & 84.5  & 84.3  & 94.8  & 81.5  & 82.1  & 81.8  & 90.4  & 83.9  \\
          & M2IO-Base-72B & 96.7  & 94.9  & 95.9  & 84.7  & 99.0  & 93.2  & 91.2  & 82.7  & 99.1  & 94.4  & 93.6  & 92.4  \\
          & M2IO-SFT-3B & 95.0  & 94.7  & 90.5  & 83.3  & 98.0  & 95.9  & 86.6  & 82.3  & 98.7  & 94.4  & 83.8  & 89.8  \\
          & M2IO-R1-3B & 95.4  & 95.4  & 93.0  & 84.1  & 98.4  & 95.9  & 88.1  & 82.7  & 99.3  & 96.8  & 85.6  & 91.1  \\
    \midrule
    \multirow{6}[2]{*}{Qwen2.5-VL-7B} & Single-Shot & 86.3  & 86.7  & 89.0  & 79.6  & 95.2  & 78.5  & 94.3  & 78.9  & 96.7  & 77.4  & 91.9  & 86.2  \\
          & Rule-Based & 68.1  & 69.7  & 76.3  & 72.2  & 60.6  & 56.6  & 57.7  & 64.3  & 64.1  & 57.1  & 59.5  & 61.2  \\
          & M2IO-Base-3B & 58.8  & 68.8  & 86.8  & 74.6  & 86.9  & 86.7  & 93.7  & 80.8  & 84.4  & 83.6  & 89.8  & 84.7  \\
          & M2IO-Base-72B & 87.8  & 89.0  & 93.7  & 81.3  & 93.2  & 89.0  & 88.0  & 80.0  & 93.4  & 88.8  & 85.7  & 87.0  \\
          & M2IO-SFT-3B & 91.6  & 90.9  & 95.3  & 82.2  & 95.2  & 89.6  & 93.0  & 81.4  & 95.7  & 84.1  & 92.2  & 88.3  \\
          & M2IO-R1-3B & 93.1  & 92.6  & 95.7  & 82.7  & 95.6  & 90.0  & 93.9  & 81.7  & 97.0  & 85.9  & 95.8  & 90.1  \\
    \bottomrule
    \end{tabular}%
  \label{tab:complete_old_data_2}%
\end{table*}%

% Table generated by Excel2LaTeX from sheet 'res_paper_full_v2'
\begin{table*}[h]
  \centering
  \setlength{\tabcolsep}{5pt} % 您可以取消注释以调整列间距
  \caption{Performance comparisons between the M2IO framework and baselines, where Inserter-SFT-3B and Inserter-R1-3B are trained on 80\% of the web-related datasets (\texttt{Wit}, \texttt{Web}, \texttt{Wiki}) and evaluated on the corresponding three MRAMG-Bench test sets (\texttt{Arxiv}, \texttt{Manual}, \texttt{Recipe}). }
    \begin{tabular}{l|cccc|ccccc|ccccc}
    \toprule
    \multicolumn{1}{c|}{\multirow{2}[2]{*}{\textbf{Strategy}}} & \multicolumn{4}{c|}{\textbf{Arxiv}} & \multicolumn{5}{c|}{\textbf{Manual}}  & \multicolumn{5}{c}{\textbf{Recipe}} \\
          & \textbf{Rec} & \textbf{F1} & \textbf{Rel} & \textbf{Ovr} & \textbf{Rec} & \textbf{F1} & \textbf{Ord} & \textbf{Rel} & \textbf{Ovr} & \textbf{Rec} & \textbf{F1} & \textbf{Ord} & \textbf{Rel} & \textbf{Ovr} \\
    \midrule
    Single-Shot & 78.3  & 68.4  & 89.5  & 74.4  & 38.6  & 37.2  & 24.5  & 75.5  & 54.5  & 50.7  & 46.4  & 33.7  & 77.3  & 59.3 \\
    Rule-Based & 64.5  & 38.3  & 52.7  & 57.6  & 35.7  & 37.6  & 23.0  & 83.6  & 55.9  & 38.4  & 50.0  & 29.3  & 79.3  & 59.5 \\
    M2IO-Base-3B & 77.3  & 53.8  & 92.3  & 71.4  & 32.1  & 31.5  & 22.1  & 84.1  & 54.6  & 57.3  & 46.8  & 37.2  & 83.8  & 61.4 \\
    M2IO-Base-72B & 85.5  & 68.5  & 96.9  & 76.2  & 50.9  & 45.6  & 35.4  & 90.9  & 61.4  & 68.0  & 55.4  & 42.4  & 89.6  & 65.3 \\
    M2IO-SFT-3B & 78.0  & 57.4  & 88.8  & 71.4  & 36.6  & 27.9  & 15.7  & 61.3  & 48.0  & 42.5  & 30.4  & 15.3  & 54.5  & 47.8 \\
    M2IO-R1-3B & 80.6  & 72.0  & 97.4  & 77.2  & 42.3  & 40.0  & 29.1  & 85.7  & 58.0  & 58.8  & 51.2  & 31.1  & 84.5  & 61.6 \\
    \bottomrule
    \end{tabular}%
  \label{tab:complete_new_data_1}%
\end{table*}%

% Table generated by Excel2LaTeX from sheet 'res_paper_full_v2'
\begin{table*}[h]
  \centering
  \caption{Performance comparisons between the M2IO framework and baselines, where Inserter-SFT-3B and Inserter-R1-3B are trained on 80\% of the web-related datasets (\texttt{Wit}, \texttt{Web}, \texttt{Wiki}) and evaluated on the corresponding three MRAMG-Bench test sets (\texttt{Web}, \texttt{Wiki}, \texttt{Wit}). }
    \begin{tabular}{l|cccc|cccc|cccc}
    \toprule
    \multicolumn{1}{c|}{\multirow{2}[2]{*}{\textbf{Strategy}}} & \multicolumn{4}{c|}{\textbf{Web}} & \multicolumn{4}{c|}{\textbf{Wiki}} & \multicolumn{4}{c}{\textbf{Wit}} \\
          & \textbf{Rec} & \textbf{F1} & \textbf{Rel} & \textbf{Ovr} & \textbf{Rec} & \textbf{F1} & \textbf{Rel} & \textbf{Ovr} & \textbf{Rec} & \textbf{F1} & \textbf{Rel} & \textbf{Ovr} \\
    \midrule
    Single-Shot & 88.4  & 90.8  & 95.0  & 83.1  & 72.3  & 72.0  & 62.4  & 70.9  & 85.1  & 84.6  & 75.0  & 76.9  \\
    Rule-Based & 48.3  & 35.7  & 45.8  & 57.1  & 55.5  & 28.7  & 32.8  & 52.7  & 54.6  & 21.2  & 32.6  & 50.4  \\
    M2IO-Base-3B & 59.6  & 69.0  & 87.8  & 75.9  & 83.2  & 82.8  & 94.1  & 81.5  & 82.6  & 82.1  & 96.5  & 81.6  \\
    M2IO-Base-72B & 91.4  & 92.4  & 94.4  & 83.4  & 98.0  & 96.2  & 93.7  & 84.7  & 96.7  & 95.6  & 94.1  & 84.4  \\
    M2IO-SFT-3B & 88.4  & 88.9  & 90.2  & 81.5  & 97.0  & 93.1  & 88.9  & 82.8  & 89.3  & 83.8  & 80.2  & 77.9  \\
    M2IO-R1-3B & 94.0  & 94.0  & 94.6  & 83.8  & 99.0  & 97.7  & 95.1  & 85.5  & 96.7  & 96.1  & 96.0  & 85.0  \\
    \bottomrule
    \end{tabular}%
  \label{tab:complete_new_data_2}%
\end{table*}%

% Table generated by Excel2LaTeX from sheet 'res_paper'
% Table generated by Excel2LaTeX from sheet 'res_paper'
% Table generated by Excel2LaTeX from sheet 'res_paper'
\begin{table*}[h]
  \centering

  \caption{Performance comparisons of different strategies on the FTII-Bench datasets.}
    \begin{tabular}{l|cc|cc|cc}
    
    \toprule
    \multicolumn{1}{c|}{\multirow{2}[2]{*}{\textbf{Strategy}}} & \multicolumn{2}{c|}{\textbf{FTII-1}} & \multicolumn{2}{c|}{\textbf{FTII-2}} & \multicolumn{2}{c}{\textbf{FTII-3}} \\
          & \textbf{F1} & \textbf{Pos} & \textbf{F1} & \textbf{Pos} & \textbf{F1} & \textbf{Pos} \\
    \midrule
    Rule-Based & 29.6  & 29.5  & 28.9  & 28.0  & 30.8  & 29.0  \\
    M2IO-Base-3B & 30.2  & 14.8  & 26.9  & 12.9  & 30.0  & 15.6  \\
    M2IO-Base-72B & 64.3  & 36.5  & 66.8  & 36.5  & 63.4  & 32.6  \\
    M2IO-SFT-3B & 46.4  & 28.4  & 48.7  & 29.9  & 43.8  & 28.0  \\
    M2IO-R1-3B & 53.7  & 30.6  & 51.2  & 32.9  & 53.2  & 31.8  \\
    \bottomrule
    \end{tabular}%
  \label{tab:ftii_results}%
\end{table*}%

\subsubsection{Complete Experiment Results.} We present the full evaluation results on all six datasets of MRAMG-Bench in Tables~\ref{tab:complete_old_data_1}, \ref{tab:complete_old_data_2} (trained on the M2IO-Inserter dataset) and Tables~\ref{tab:complete_new_data_1}, \ref{tab:complete_new_data_2} (training on web-style data only). And the evaluation results on all three FTII-Bench datasets are shown in Table~\ref{tab:ftii_results}.

\subsubsection{Ablation Study.}
We conduct an ablation study to evaluate the impact of the hyperparameter $\alpha$ within the reward function $\mathcal{R}_{\text{answer}}$ during the training of the Inserter-R1-3B model. The value of $\alpha$ is critical for balancing the trade-off between image recall and positional accuracy. Specifically, we conduct a hyperparameter search for $\alpha$, with the search space defined as:
$\alpha \in \{0.0, 0.2, 0.4, 0.5, 0.6, 0.8, 1.0\}.$

As shown in Table \ref{ablation_1} and Table \ref{ablation_2}, we compare the performance of the M2IO-R1-3B method across seven different hyperparameter configurations and against several baselines, with GPT-4o used solely as the text generator. Notably, the $\alpha = 0.8$ exhibits superior and consistently well-balanced performance across diverse datasets, highlighting its effectiveness in jointly optimizing image insertion quality—by accurately selecting relevant visual content, excluding semantically irrelevant distractors, and inserting images in an order that faithfully reflects the underlying textual structure and intent.

Except for the extreme cases of $\alpha = 0$ and $\alpha = 1$, M2IO-R1-3B consistently meets or exceeds all baselines across evaluation metrics on six datasets, demonstrating its robustness and generalizability. For instance, on the Manual dataset, M2IO-R1-3B (Inserter-R1-3B with $\alpha = 0.4$) achieves an F1 score of 43.2, significantly outperforming the M2IO-SFT-3B, which attains an F1 score of 38.6. On the Arxiv dataset, M2IO-R1-3B (Inserter-R1-3B with $\alpha = 0.5$) reaches an F1 score of 70.4, even exceeding that of M2IO-Base-72B, which achieves 69.3. These results further highlight the robustness and effectiveness of our method in achieving accurate interleaved image-text responses.

\subsection{Case Analysis}
\begin{figure*}[h]
    \centering    \includegraphics[width=0.9\linewidth]{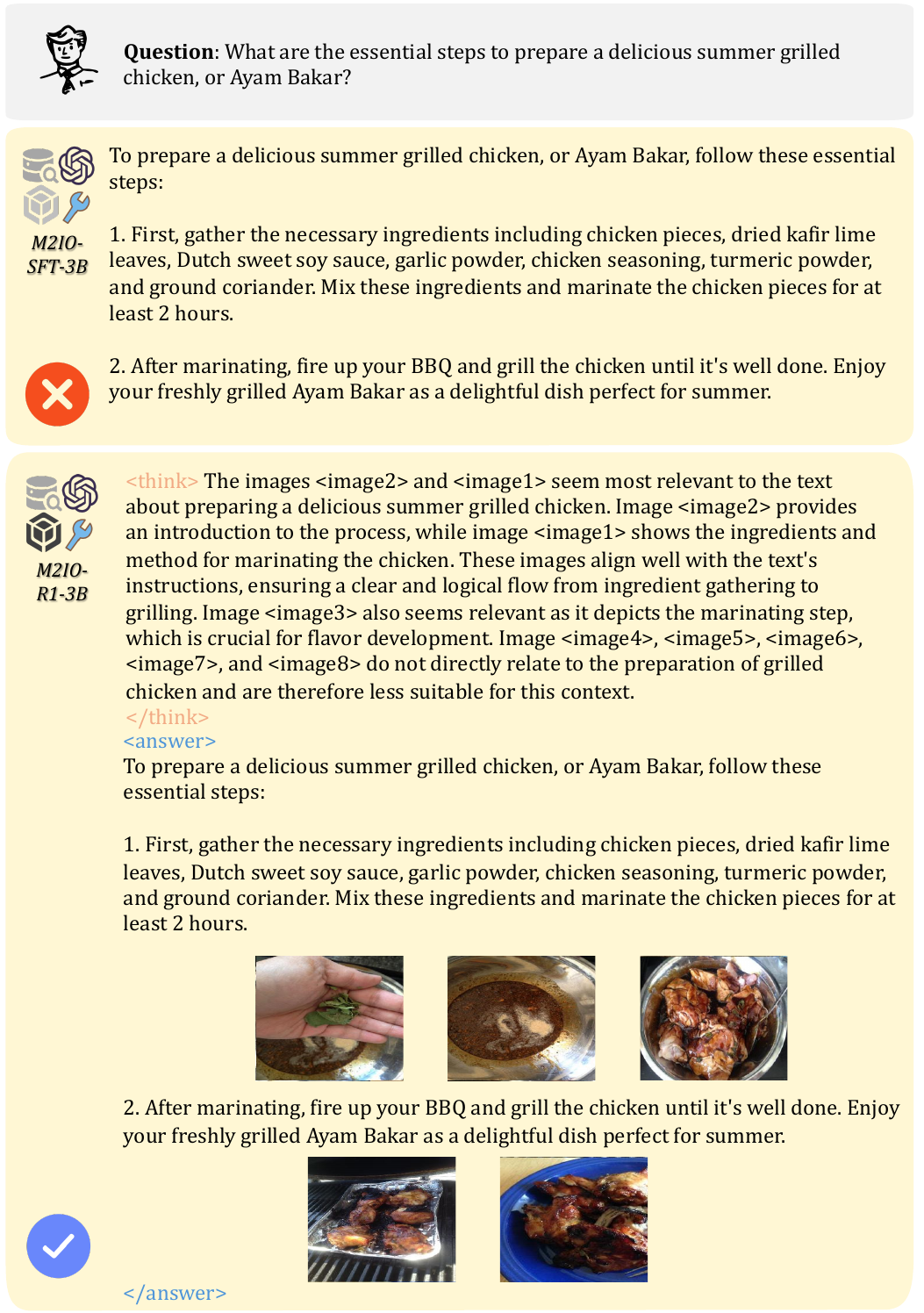}
   \caption{ 
    %A case of comparison between the plain text output of M2IO-SFT-3B and the more expected multi-modal output of M2IO-R1-3B.
    A case comparing the plain text output of M2IO-SFT-3B, which omits the expected illustrations, with the comprehensive multimodal output of M2IO-R1-3B.
    }
    \label{fig:case1}
\end{figure*}

\begin{figure*}[h]
    \centering    \includegraphics[width=0.75\linewidth]{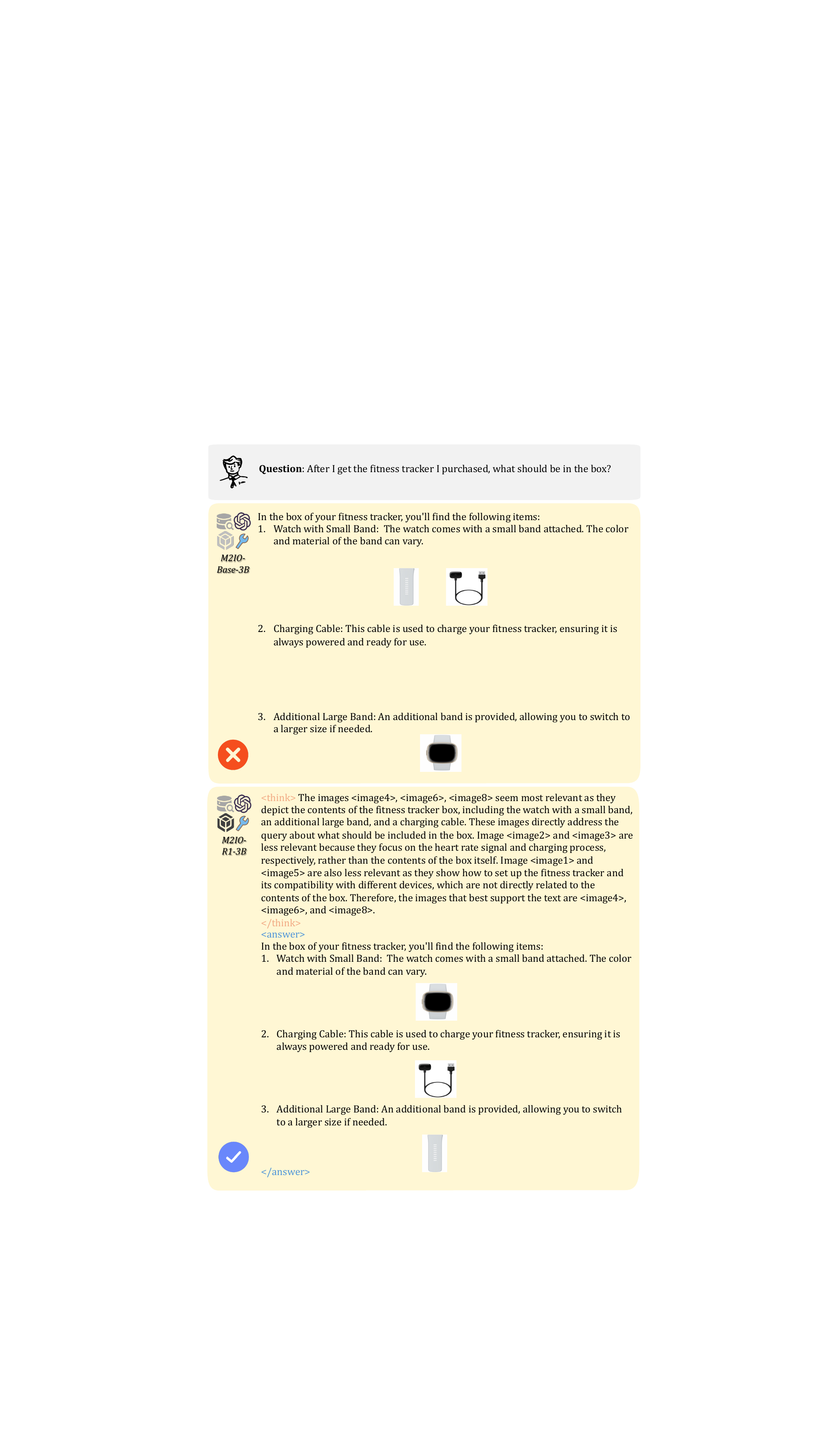}
   \caption{ 
    %A case of comparison between the confusing multimodal output with inappropriate images placement of M2IO-Base-3B and the more expected multimodal output of M2IO-R1-3B with accurate images placement.
    A case contrasting the multimodal output of M2IO-Base-3B, where images are placed inappropriately, with the expected multimodal output of M2IO-R1-3B, which features accurate image placement.
    }
    \label{fig:case2}
\end{figure*}

\begin{figure*}[h]
    \centering    \includegraphics[width=0.95\linewidth]{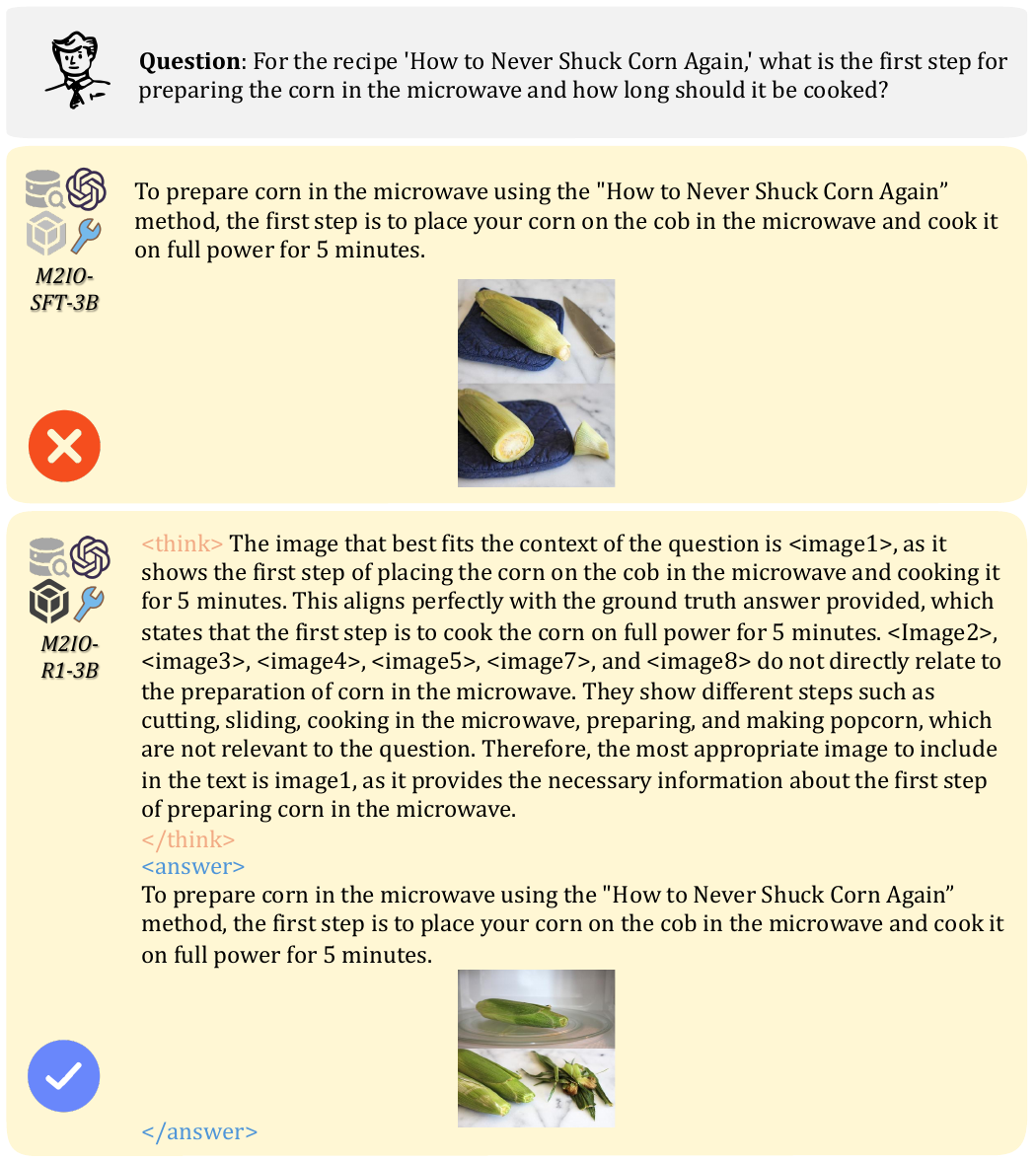}
   \caption{ 
    %A case of comparison between the confusing multimodal output with incorrect images selection of M2IO-SFT-3B and the more expected multi-modal output of M2IO-R1-3B with accurate images selection.
    A case contrasting the multimodal output of M2IO-SFT-3B, which exhibits incorrect image selection, with the expected multimodal output of M2IO-R1-3B, featuring accurate image selection.
    }
    \label{fig:case3}
\end{figure*}

\begin{figure*}[h]
    \centering    \includegraphics[width=0.65\linewidth]{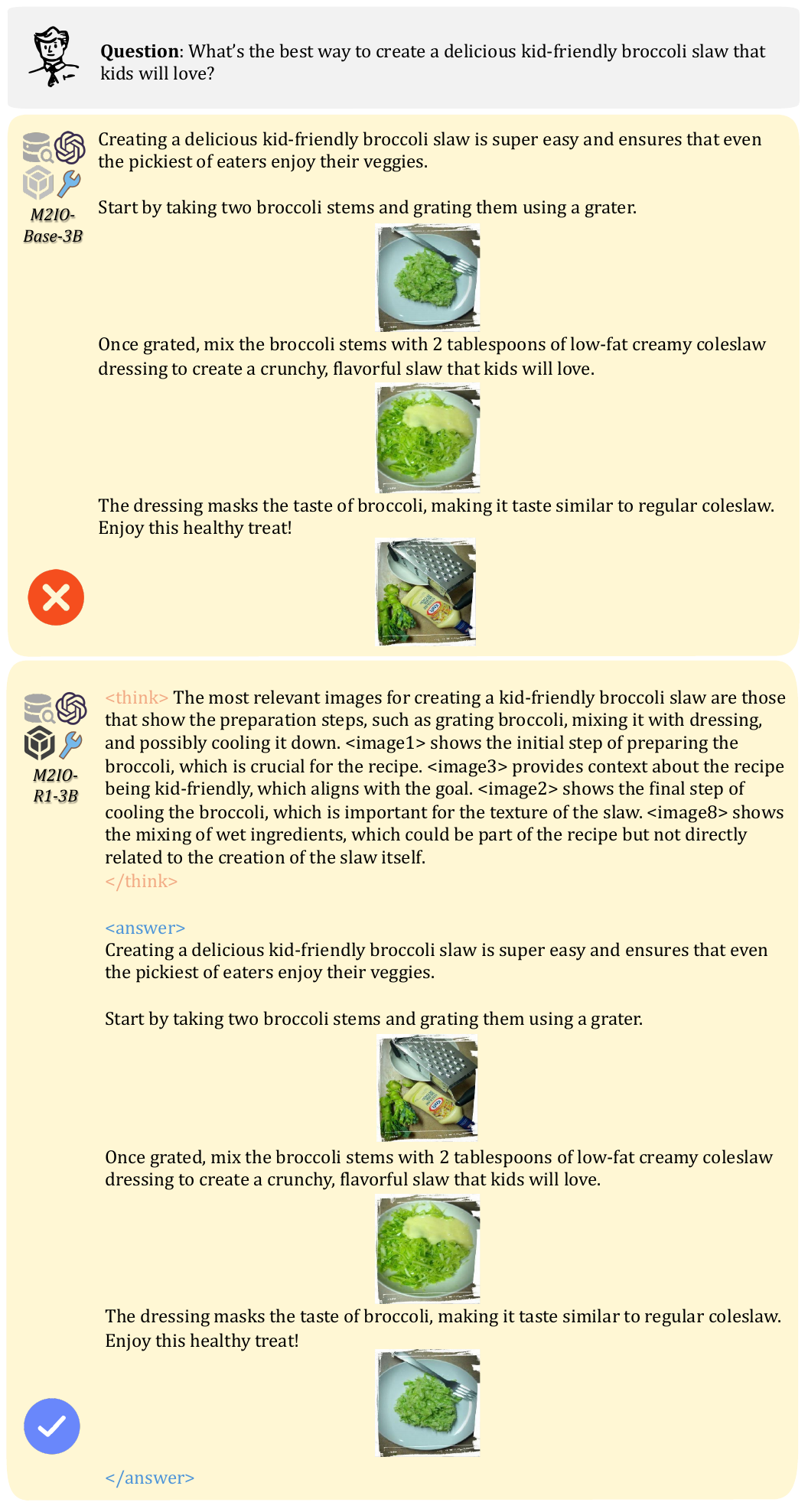}
   \caption{ 
    %A case of comparison between the confusing multimodal output with incorrect images order of M2IO-Base-3B and the more expected multi-modal output of M2IO-R1-3B with accurate images order.
    A case comparing the multimodal output of M2IO-Base-3B, which exhibits incorrect image ordering, with the expected multimodal output of M2IO-R1-3B, which ensures accurate image ordering.
    }
    \label{fig:case4}
\end{figure*}

In this section, we provide specific cases to showcase the output quality differences of between our approach with other baselines.
\begin{itemize}
    \item M2IO-R1-3B delivers rich and precise multimodal results, whereas M2IO-SFT-3B generates only plain text, omitting images entirely (Figure \ref{fig:case1}).
    \item With M2IO-R1-3B, images are accurately integrated into the output, in sharp contrast to M2IO-Base-3B, which places images in inappropriate locations (Figure \ref{fig:case2}).
    \item The outputs of M2IO-R1-3B showcase carefully chosen and contextually relevant images, while M2IO-SFT-3B incorrectly selects images that do not align with the content (Figure \ref{fig:case3}).
    \item M2IO-R1-3B ensures coherent and logically ordered image presentation, unlike M2IO-Base-3B, which produces outputs with images arranged in the wrong sequence (Figure \ref{fig:case4}).
\end{itemize}

These qualitative comparisons underscore the superiority of our approach. M2IO-R1-3B consistently generates coherent multimodal content by ensuring not only the contextual relevance of images but also their proper placement and logical ordering. In contrast, the baseline models either fail to produce multimodal outputs entirely or exhibit significant deficiencies in image selection and structural arrangement, highlighting the advanced capabilities of our method

% Table generated by Excel2LaTeX from sheet 'ablation study table'

% Table generated by Excel2LaTeX from sheet 'ablation study table'
\begin{table*}[t]
  \centering
\setlength{\tabcolsep}{3pt}   
  
  \caption{Ablation study on the hyperparameter $\alpha$ in the reward function $\mathcal{R}_{\text{answer}}$ during Inserter-R1-3B training. These models are trained on Full-source datasets and evaluated on the corresponding three MRAMG-Bench test sets (\texttt{Arxiv}, \texttt{Manual}, \texttt{Recipe}).}

    \begin{tabular}{l|c|cc|ccc|ccc}
 \toprule
    \multicolumn{1}{c|}{\multirow{2}[1]{*}{\textbf{Strategy}}} & \textbf{Answer Reward} & \multicolumn{2}{c|}{\textbf{Arxiv}} & \multicolumn{3}{c|}{\textbf{Manual}} & \multicolumn{3}{c}{\textbf{Recipe}} \\
          & \boldmath{}\textbf{$\alpha$}\unboldmath{} & \textbf{Rec} & \textbf{F1} & \textbf{Rec} & \textbf{F1} & \textbf{Ord} & \textbf{Rec} & \textbf{F1} & \textbf{Ord} \\
    \midrule
    \multirow{7}[1]{*}{M2IO-R1-3B} & 0     & 66.5  & 67.4  & 25.3  & 31.4  & 23.6  & 44.3  & 48.4  & 43.1  \\
          & 0.2   & 76.7  & 69.7  & 50.9  & 46.0  & 36.7  & 69.3  & 55.7  & 42.9  \\
          & 0.4   & 82.0  & 69.0  & 49.1  & 43.2  & 32.1  & 66.0  & 55.7  & 41.7  \\
          & 0.5   & 83.7  & 70.4  & 52.3  & 44.5  & 35.0  & 73.4  & 56.3  & 43.3  \\
          & 0.6   & 83.7  & 69.9  & 52.1  & 45.8  & 37.5  & 69.1  & 53.7  & 40.4  \\
          & 0.8   & 84.2  & 68.4  & 52.6  & 46.5  & 39.4  & 68.2  & 56.4  & 44.3  \\
          & 1     & 76.0  & 53.1  & 51.3  & 42.2  & 29.7  & 73.2  & 50.4  & 33.8  \\
          \midrule
    Single-Shot & -     & 80.1  & 69.1  & 39.7  & 38.0  & 25.6  & 50.1  & 45.9  & 33.0  \\
    Rule-Based  & -     & 65.7  & 57.5  & 44.8  & 44.3  & 32.4  & 67.7  & 52.0  & 39.0  \\
    M2IO-Base-3B & -     & 76.0  & 55.0  & 32.7  & 31.9  & 23.1  & 57.8  & 47.4  & 37.8  \\
    M2IO-Base-72B & -     & 83.7  & 69.3  & 52.7  & 47.3  & 38.6  & 67.9  & 55.3  & 42.2  \\
    M2IO-SFT-3B & -     & 75.7  & 64.6  & 40.9  & 38.6  & 30.3  & 62.7  & 57.1  & 45.1  \\
    \bottomrule
    \end{tabular}%
  \label{ablation_1}
\end{table*}%

% Table generated by Excel2LaTeX from sheet 'ablation study table'
\begin{table*}[h]
  \centering

  \caption{Ablation study on the hyperparameter $\alpha$ in the reward function $\mathcal{R}_{\text{answer}}$ during Inserter-R1-3B training. These models are trained on Full-source datasets and evaluated on the corresponding three MRAMG-Bench test sets (\texttt{Web}, \texttt{Wiki}, \texttt{Wit}).}
    \begin{tabular}{l|c|cc|cc|cc}
    \toprule
    \multicolumn{1}{c|}{\multirow{2}[3]{*}{\textbf{Strategy}}} & \textbf{Answer Reward} & \multicolumn{2}{c|}{\textbf{Web}} & \multicolumn{2}{c|}{\textbf{Wiki}} & \multicolumn{2}{c}{\textbf{Wit}} \\
         & \boldmath{}\textbf{$\alpha$}\unboldmath{} & \textbf{Rec} & \textbf{F1} & \textbf{Rec} & \textbf{F1} & \textbf{Rec} & \textbf{F1} \\
    \midrule
    \multirow{7}[1]{*}{M2IO-R1-3B} & 0     & 74.3  & 81.8  & 98.4  & 98.4 & 97.3  & 97.2  \\
          & 0.2   & 89.9  & 91.4  & 97.6  & 97.3  & 97.7  & 97.1  \\
          & 0.4   & 88.2  & 90.0  & 97.6  & 97.3  & 97.0  & 96.2  \\
          & 0.5   & 93.5  & 93.2  & 98.4  & 97.3  & 98.0  & 96.4  \\
          & 0.6   & 92.0  & 92.9  & 97.2  & 96.5  & 97.3  & 95.6  \\
          & 0.8   & 93.3  & 93.3  & 98.4  & 97.5  & 99.0  & 97.3  \\
          & 1     & 96.1  & 95.4  & 98.0  & 90.0  & 99.3  & 85.3  \\
          \midrule
    Single-Shot & -     & 90.5  & 90.5  & 72.9  & 72.8  & 85.1  & 84.7  \\
    Rule-Based  & -     & 59.7  & 59.7  & 53.0  & 52.9  & 61.1  & 60.8  \\
    M2IO-Base-3B & -     & 70.7  & 70.7  & 82.9  & 82.7  & 85.4  & 84.9  \\
    M2IO-Base-72B & -     & 92.8  & 93.3  & 96.4  & 95.3  & 96.4  & 95.0  \\
    M2IO-SFT-3B & -     & 92.1  & 92.1  & 98.4  & 97.7  & 98.3  & 94.9  \\
    \bottomrule
    \end{tabular}%
   \label{ablation_2}
\end{table*}%

\newpage

